\documentclass[acmsmall]{acmart}

\AtBeginDocument{%
  }

\setcopyright{acmlicensed}
\acmDOI{10.1145/3643753}
\acmYear{2024}
\copyrightyear{2024}
\acmSubmissionID{fse24main-p437-p}
\acmJournal{PACMSE}
\acmVolume{1}
\acmNumber{FSE}
\acmArticle{27}
\acmMonth{7}
\received{2023-09-20}
\received[accepted]{2024-01-23}

\usepackage{array}
\usepackage{amsmath,amsfonts}
\usepackage[noend]{algpseudocode}
\usepackage{graphicx}
\usepackage{textcomp}
\usepackage{float}
\usepackage{listings}
\usepackage{xspace}
\usepackage{multirow}
\usepackage{amsthm}

\usepackage{balance}
\usepackage{algorithm}
\usepackage{algpseudocode}

\usepackage[skins]{tcolorbox}
\usepackage{xcolor,pifont}

\usepackage{graphicx,stackengine}

\newcommand*\colourcheck[1]{%
	\expandafter\newcommand\csname #1check\endcsname{\textcolor{#1}{\ding{52}}}%
}
\colourcheck{blue}
\colourcheck{green}
\colourcheck{red}

\newtcolorbox{boxB}[2][]{%
  enhanced,colback=white,colframe=black,coltitle=black,
  sharp corners,
  toprule=1.0pt,
  rightrule=0.3pt,
  leftrule=0pt,
  bottomrule=0pt,
  fonttitle=\itshape\scshape\large,
  left=0pt,right=5pt,top=5pt,bottom=3pt,
  attach boxed title to top right={yshift=-0.3\baselineskip-0.4pt,xshift=-5mm},
  boxed title style={tile,size=minimal,left=0.2mm,right=0.5mm,
    colback=white,before upper=\strut},
  title=#2,#1
}

\newcommand{\tool}{\textsc{SlimCode}\xspace}
\newcommand{\toolbase}{\textsc{DietCode}\xspace}

\newboolean{showcomments}
\setboolean{showcomments}{true}
\ifthenelse{\boolean{showcomments}}
 { \newcommand{\mynote}[2]{
      \fbox{\bfseries\sffamily\scriptsize#1}
        {\small$\blacktriangleright$\textsf{\emph{#2}}$\blacktriangleleft$}}}
        { \newcommand{\mynote}[2]{}}

\newcommand{\todoc}[2]{{\textcolor{#1} {\textbf{#2}}}}

\newcommand{\todored}[1]{\todoc{red}{\textbf{#1}}}

\newcommand{\david}[1]{\mynote{David}{\todored{#1}}}

\usepackage{tikz}
\newcommand*\circled[1]{\tikz[baseline=(char.base)]{
            \node[shape=circle,draw,inner sep=1.5pt] (char) {#1};}}

\newcommand{\intuition}[1]{
\begin{tcolorbox}[colback=white,boxrule=1pt,top=0pt,bottom=0pt,left=1pt,right=2pt,top=2pt,bottom=2pt]%[tile,size=fbox,boxsep=2mm,boxrule=0pt,top=0pt,bottom=0pt,borderline={0.5mm}{0pt}{black!70!white},colback=black!5!white]
\em #1
\end{tcolorbox}
}

\newcolumntype{L}[1]{>{\raggedright\arraybackslash}p{#1}}

\newcommand{\code}[1]{{\footnotesize\texttt{#1}}}
\usepackage{amsthm}
 \definecolor{dkgreen}{rgb}{0,0.6,0}
\definecolor{gray}{rgb}{0.5,0.5,0.5}
\definecolor{mauve}{rgb}{0.58,0,0.82}
\begin{document}

%\title{The Name of the Title Is Hope}

\title{Natural Is the Best: Model-Agnostic Code Simplification for Pre-trained Large Language Models}

\author{Yan Wang}
\authornote{Both authors contributed equally to the paper}
\orcid{0000-0002-9876-5823}
\affiliation{%
  \institution{Central University of Finance and Economics}
  \city{Beijing}
  \country{China}
}
\email{dayanking@gmail.com}

\author{Xiaoning Li}
\authornotemark[1]
\orcid{0009-0004-6845-2721}
\affiliation{%
  \institution{Central University of Finance and Economics}
  \city{Beijing}
  \country{China}
}
\email{2022212378@email.cufe.edu.cn}

\author{Tien N. Nguyen}
\orcid{0009-0006-7962-6090}
\affiliation{%
  \institution{University of Texas at Dallas}
  \city{Dallas}
  \country{USA}
}
\email{tien.n.nguyen@utdallas.edu}

\author{Shaohua Wang}
\authornote{Corresponding Author}
\orcid{0000-0001-5777-7759}
\affiliation{%
  \institution{Central University of Finance and Economics}
  \city{Beijing}
  \country{China}
}
\email{davidshwang@ieee.org}

\author{Chao Ni}
\orcid{0000-0002-2906-0598}
\affiliation{%
  \institution{Zhejiang University}
  \city{Hang Zhou}
  \country{China}
}
\email{chaoni@zju.edu.cn}

\author{Ling Ding}
\orcid{0009-0005-8189-2040}
\affiliation{%
  \institution{Central University of Finance and Economics}
  \city{Beijing}
  \country{China}
}
\email{2021312380@email.cufe.edu.cn}

\renewcommand{\shortauthors}{Wang, Li, Nguyen, Wang, Ni, Ding}

\begin{abstract}
Pre-trained Large Language Models (LLM) have achieved remarkable successes in several domains. However, code-oriented LLMs are often heavy in computational complexity, and
quadratically with the length of the input code sequence. Toward simplifying the input program of an LLM, the state-of-the-art approach has the strategies to filter the input code tokens based on the attention scores given by the LLM. The decision to simplify the
input program should not rely on the attention patterns of an LLM, as these patterns are influenced by both the
model architecture and the pre-training dataset. 
Since the model and dataset are part of the solution domain, not the problem domain
where the input program belongs, the outcome may differ when the model~is pre-trained on a different dataset. We propose {\tool}, a model-agnostic code simplification solution for LLMs that depends on the nature of input code tokens. As an empirical study on the LLMs including CodeBERT, CodeT5, and GPT-4 for two main tasks: code search and summarization, we reported that 1) the removal ratio of code has a linear-like relation with the saving ratio on training time, 2) the impact of categorized tokens on code simplification can vary significantly, 3) the impact of categorized tokens on code simplification is task-specific but model-agnostic, and 4) the above findings hold for the paradigm–prompt engineering and interactive in-context learning. The empirical results showed that {\tool} can improve the state-of-the-art technique by 9.46\% and 5.15\% in terms of MRR and BLEU score on code search and summarization, respectively. More importantly, {\tool} is 133 times faster than the state-of-the-art approach. Additionally, {\tool} can reduce the cost of invoking GPT-4 by up to 24\% per API query, while still producing comparable results to those with the original code. With this result, we call for a new direction on code-based, model-agnostic code simplification solutions to further empower LLMs.
\end{abstract}
%\david{fill up the xx\%}





%\keywords{Do, Not, Us, This, Code, Put, the, Correct, Terms, for, Your, Paper}





\maketitle

\section{Introduction}
\label{sec:intro}
Pre-trained Large Language Models (PLLM) have emerged as powerful tools that can significantly impact the way software code is written, reviewed, and optimized, making them invaluable resources for programmers. They offer developers the ability to leverage pre-trained knowledge and tap into vast code repositories, enabling faster development cycles and reducing the time spent on repetitive or mundane coding tasks. The abilities to summarize the code in concise textual descriptions or to search for desired code are also valuable to developers in code comprehension. However, while these models offer substantial benefits, their adoption also presents multiple challenges. First, the fine-tuning and prediction for downstream tasks (e.g., code summarization or code search) with pre-trained large language models can be time-consuming and prohibitively costly, and eventually becomes practically infeasible. For example, an article in CNBC news reported~\cite{Latitude} that the cost to develop and maintain the software can be extremely high with PLLMs. The company in the article was estimated to be spending lots of time and nearly \$200,000 a month on PLLMs in order to keep up with the millions of user queries it needed to process each day. Second, due to the need for massive computational resources to train, fine-tune and predict these models effectively, they often place the limit on the number of input words or code tokens. The limit for the number of tokens to be fed to CodeBERT when run locally is 512 tokens. The word position embedding matrix for GPT-2 is (1,024x768) and for GPT-3 is (2,048x128). Other large language models, (e.g., CodeT5~\cite{wang-etal-2021-codet5}, CodeGen~\cite{codegen}, GPT-4~\cite{ChatGPT}), are heavy in computational complexity and quadratically with the length of the input code sequence. The key question is \textit{``can we reduce the cost of applying PLLM in code-related tasks by simplifying its input, while maintaining the same performance?''}

Toward overcoming those challenges, DietCode~\cite{dietcode-fse22} simplifies the input program of CodeBERT~\cite{codebert-emnlp20} with three strategies, namely, word dropout, frequency filtering, and an attention-based strategy which retains the statements and tokens that receive the {\em most attention weights} during pre-training. The strategies were chosen according to a preliminary empirical analysis from the authors, reporting that 
CodeBERT pays more attention to certain types of tokens and statements. Experimental results on a dataset with two downstream tasks show that DietCode provides comparable results to CodeBERT with 40\% less computational cost in fine-tuning and testing.

However, we conjecture that the {\em decision on simplifying the input program should depend on the nature of the source code}, rather than on what input tokens the model pays more or less attention. 
%The reason is that the result on what input tokens having drawn more or less attention from CodeBERT depends on both the model itself and the dataset on which it was pre-trained on. The CodeBERT model and the pre-training dataset belong to the solution domain, and the result of that solution might vary when it is pre-trained in a different dataset.
Let us elaborate this point further. The decision to simplify the input program should not rely on the attention patterns of the CodeBERT model, as these patterns are influenced by both the model architecture and the pre-training dataset. Since the model and dataset are part of the solution domain, not the problem domain where the input program belongs, the outcome may differ when the model is pre-trained on different datasets.
The rationale is that the attention patterns can be influenced by various factors, including the specific model architecture and the training data. Different datasets may exhibit different linguistic patterns, programming styles, or coding conventions, which can affect the attention patterns learned by the model during pre-training. Consequently, relying solely on attention patterns to determine the importance of input tokens may not yield consistent results across different datasets or model~variations.

To ensure robustness and generalizability, it is advisable to consider a broader range of factors when making decisions about program simplification. 
We hypothesize that those factors could include the program's lexical tokens, syntactic structures, control flows, program dependencies, and the desired outcome of the simplification process. Those factors can be valuable in guiding the simplification process, especially when dealing with potential variations introduced by different pre-training datasets. Keeping the above hypothesis in mind, we conducted our empirical study with the following research questions:

\begin{enumerate}
    \item {\bf RQ-1: What is the impact of randomly removing code tokens on the performance of PLLM?} Our objective is to explore the relationship between the proportion of code removed and the efficiency of code summarization and code search of pre-trained large language models. To achieve this, we use random tokens with various simplification ratios to assess the impact on the training time and performance of PLLMs for downstream tasks.

    \item {\bf RQ-2: What is the impact of removing lexical tokens on the performance of PLLM?} We remove simple type tokens from the code, which are the most basic elements in the program and account for a large proportion. 
    
    %Some of them are usually considered as the main objects of eliminated tokens.
    
    \item {\bf RQ-3: What is the impact of removing syntactical tokens on the performance of PLLM?} In order to examine the impact of tokens in key syntax structures on code simplification, we use Abstract Syntax Tree (AST) to select tokens to be removed, including the tokens in control structures, method invocations and signatures.

    \item {\bf RQ-4: What is the impact of removing semantically non-essential tokens on the performance of PLLM?} We aim to identify the token types deemed to be semantically non-essential for removal, i.e., those with minimal impact on performance. Given the prevalent use of Program Dependence Graph (PDG) in representing source code semantics, e.g., in semantic code clone detection, we aim to remove the tokens absent in the PDG and evaluate the results on code-related downstream tasks.

\end{enumerate}

In RQ-1, we evaluate the impact of randomly removing tokens from code on the performance of PLLMs for two tasks: {\em code search and code summarization}. We use five different levels of removal ratio and training time, as well as two different performance metrics, to measure the relations between removal ratio and the efficiency and effectiveness of the models CodeBERT and CodeT5. In RQ-(2-4), we systematically remove tokens by categories and structures to analyze the impact of different token types on PLLMs. We categorize tokens into six categories that reflect the three different natural levels of code, and use the tools  AST-maven~\cite{ast-maven} and JAST2DyPDG~\cite{pdg-tool}, to generate ASTs and PDGs for source code, respectively. 

The key findings of our empirical study are highlighted:

\begin{itemize}
    \item {\bf Key Finding 1 (RQ-1)}: The code simplification ratio has a linear-like relationship with the saving ratio on training time. For example, as 50\% of the code was removed, the training time was also reduced by 51.5\% and 31.76\% on code search and summarization with CodeBERT. 
    
    \item {\bf Key Finding 2 (RQ-(2-4))}: The impact of token categories on code simplification varies significantly. E.g., removing symbol tokens that account for 51.38\% of total code tokens results only causes a 2.8\% decrease in CodeBERT's performance on code search. Meanwhile, removing variables that account for 15.48\% of total tokens leads to a 12.52\% decrease in CodeBERT's performance.
    
    \item {\bf Key Finding 3 (RQ-(2-4))}: The impact of categorized tokens on code simplification is task-specific but model-agnostic. In the code summarization/search for both CodeBERT and CodeT5, the impact of the method signature/variables removal is the most significant on the models' performance, while the impact of symbol tokens removal is the least significant.

\end{itemize}

%    \item {\bf RQ-5: Can we propose an effective code simplification method?} Considering the aforementioned impacts, it is necessary to offer users a lightweight code simplification method that leverages the inherent properties of code. This will enable efficient and effective code simplification, regardless of the underlying model being used.

%    \item {\bf RQ-6: Can the findings in the above RQ-s be applied to the pre-train, prompt and prediction paradigm?} The above research questions are based on the pre-train and fine-tune paradigm. Currently, more PLLM work on the pre-train, prompt, and prediction paradigm. The objective of this study is to explore if our conclusions and method have the ability of paradigm transfer.

The above results on the impacts of different types of code tokens motivate us to develop a code simplification method that leverages the inherent properties of source code. This will enable efficient and effective code simplification, regardless of the underlying LLMs. We firstly model the problem of code simplification as 0-1 knapsack problem and propose {\tool}, a novel method of code simplification based on the above empirical findings. {\tool} is applied to two classical downstream tasks (classification task and generative task) and we measure the performance. 
The empirical results in Section~\ref{slimcoderesults} show that \tool can improve the state-of-the-art
technique, DietCode, by 9.46\% and 5.15\% in terms of MRR and BLEU score on code search and summarization,
respectively. More importantly, \tool is 133 times faster than DietCode.

%Our empirical results show that, compared to the state-of-the-art \toolbase,  {\tool} achieves average improvements of \textcolor{blue}{{\bf5.97\%}} and \textcolor{blue}{{\bf4.23\%}} for both tasks, while also demonstrating \textcolor{blue}{{\bf 133 times}} faster performance on the same datasets.

%Our empirical results show that {\tool} is up to \textcolor{red}{{\bf 133 times}} faster than the state-of-the-art $Dietcode$ with \textcolor{red}{{\bf significantly better performance at various token removal ratios}} on the same datasets. 

Additionally, we also found that the results reported in RQ-(1-4) are valid for the paradigm of pre-training, prompt engineering, and interactive in-context learning with GPT-4 (Section~\ref{chatgptresults}). For code search, \tool can reduce the inference time of GPT-4 by 27\% and the cost of invoking GPT-4 by 24\% per API query, with a 2.3\% improvement of precision compared to results with the original code. In this paper, we make the following key contributions:

\begin{itemize}

  \item {\bf The state-of-the-art knowledge}. Our study is the first to leverage the nature of code to simplify the input code for pre-trained large language models and GPT-like models.
  
  \item {\bf {\tool}: a new code simplification approach}. We quantitatively analyzed the impacts of different types of code tokens and proposed {\tool}, a model-agnostic code simplification method for PLLMs. Our code and data are available at https://github.com/gksajy/slimcode.
  %based on the knapsack optimization.   
  %\david{I changed the tool's name to SlimCode}
  %\item In both paradigms, \tool can significantly reduce the running time by properly removing tokens while maintaining acceptable effects. Its performance can be comparable to, or even better than, the state-of-the-art method Dietcode.              
\end{itemize}

\section{Token category and removal}
\label{sec:simplification}

Large language models based on the Transformer architecture are aimed to learn the {\em lexical, syntactic, and semantic} knowledge of programming languages after pre-training, and map them to token vectors. Each token carries varying amounts of information at different levels. By categorizing tokens, we identify which words carry informative "hints" for downstream tasks.

In this study, we categorize different tokens in code snippets based on their nature.
We consider six categories: {\em symbol tokens, identifiers, tokens in control structures, tokens in method invocations, tokens in method signatures, and tokens in Program Dependence Graphs (PDG)}.

The first three categories are at the lexical level, as the selection criteria considers only the inherent features of the tokens themselves. The next three categories are at the syntactic level because their selection requires considering the grammar. The final category is at the semantic level, as token selection depends on program semantics. Let us provide more details.

%Detailed information on each category is provided in the following sections.

\subsection{Lexical Level}
\label{sec:lex}

At the lexical level, we chose {\em symbol tokens} and {\em identifiers}. 

\subsubsection{Symbol Tokens}

Symbol tokens are defined as the ones including brackets, separators and operators, {\em i.e.}, \{=,+, -, *, /, \%, $++$, $--$, !, $==$, $!=$, $>$,$>=$, $<$, $<=$, $||$, \~, $<<$, $>>$, $>>>$, |, $($, $)$, \{, \}, [, ], ,, ;, ., \^,\}.  
Symbol tokens are extensively used in programming to form the fundamental building blocks in a program. They specify the delimiters and the detailed instructions to the computer in certain operations. 
%to write code that can process and manipulate data in meaningful ways. 
However, as the most basic elements, symbol tokens usually carry less high-level, syntax information and high-level semantic, downstream task-related information. Thus, removing such symbols is expected to greatly simplify code, as they make up a significant portion of code tokens.

\subsubsection{Identifiers}

Identifiers convey the intention of developers in programming tasks. Readability is a defining feature of identifiers, as they allow variables to have meaningful and descriptive names. This can improve code readability and make better understandability. Improved readability results in more "hints" for downstream tasks.
This study examines the effects of removing identifiers. 

%The following example demonstrates removing identifiers:

%This is the origin source code are listed as follows:
%\begin{lstlisting}
%public static boolean safeClose ( Closeable cb ) { 
%	if ( null != cb ) 
%		try { 
%			cb.close ( ); 
%		} catch ( IOException e ) { 
%			return false; 
%		} 
%	return true ; 
%}
%\end{lstlisting}
%The result of removing all the identifiers of %the original example are shown as follows:
%\begin{lstlisting}
%public static boolean safeClose(Closeable ) { 
%	if (null != ) try { 
%		.close(); 
%	} catch (IOException e) { 
%		return false; 
%	} 
%	return true; 
%}
%\end{lstlisting}

\subsection{Syntactic Level}
\label{sec:syntax}
At the syntactic level, we use AST to select tokens to be removed. We consider three types of tokens for removal: {\em tokens in control structures}, {\em tokens in method invocations, and tokens in method signatures}.
The tokens at the syntactic level carry more information about the code's content. They reflect the business logic behind the code to some extent and could provide more hints for solving downstream tasks. Therefore, we aim to study the impact of removing syntactic-level tokens.

%At the syntactic level (AST), we chose the structural tokens. {\em Structural tokens} are defined as the code tokens that play the syntactic roles. For example, the token \code{if} defines a conditional structure; the token \code{for} defines a repetition in the code, etc. {\bf \color{red}{The syntactic structure is crucial for program correctness, error detection, portability, and collaboration. In particular, when it comes to readability, structural tokens that carry more information about the details of the code (such as control flow and branch conditions) are more helpful than lexical-level tokens. Again, usually readability is the key for downstream tasks ands thus we should take into account how removing structure tokens would affect our work. To be specific, structure tokens here includes 'for', 'if', 'try', 'catch', 'switch', 'while', 'do while' and their judgment conditions. In the following example, 'if', ‘try’, 'catch' and their conditions are removed.}}

\subsubsection{Tokens in Control Structures}
Tokens in control structures are defined as the code tokens that are used as keywords for control-flow statements. For example, the token \code{if} defines a conditional structure; the token \code{for} defines a repetition in the code, etc. 
To be specific, the tokens in control structures include the following: \code{for}, \code{if}, \code{try}, \code{catch}, \code{switch}, \code{while}, \code{do while} and their conditions.

%In the following example, \code{if}, \code{try}, \code{catch} and their corresponding conditions are removed.

%\begin{lstlisting}
%public static boolean safeClose(Closeable cb) {           
%	(){
%    cb.close();              
%	} () {                  
%		return false;              
%	}          
%		return true;      
%}
%\end{lstlisting}

\subsubsection{Tokens in Method Signatures}

Tokens in a method signature include the ones in the signature and in the parameters of the method. 
This type tokens can be found in the \code{MethodDeclaration} and \code{Parameter} nodes in an AST.
Method signatures are also strongly related to code content, meanwhile they can provide information about the inputs and outputs of a method. 
This can make it easier for models to understand how the method works and how it can be used in their own code.

%After the method signature is removed, it becomes like the below.
%\begin{lstlisting}
%() { 
%	if ( null != cb ) 
%		try{ 
%			cb.close ( ); 
%		} catch ( IOException e ) { 
%			return false; 
%		} 
%	return true; 
%}
%\end{lstlisting}

\subsubsection{Tokens in Method Invocations}

Method invocations include calls to methods of objects as well as calls to methods within the same class.
Notice that a method invocation typically consists of method name, parameters, and return value(s), and all the information can provide potential or direct hints for the downstream tasks, such as the purpose and functionality of the method.

%The following is the result of removing the method invocation from the previous example.
%\begin{lstlisting}
%public static boolean safeClose ( Closeable cb ) { 
%	if ( null != cb ) 
%		try { 
%		} catch ( IOException e ) { 
%			return false; 
%		} 
%	return true; 
%}
%\end{lstlisting}

\subsection{Semantic Level}
\label{sec:semantic}

%At the semantic level, we chose {\em method signatures}, {\em method invocations}, and the {\em statements in a program dependence graph (PDG)}.

%\subsubsection{Removing Statements outside of PDG}

Our goal is to identify the token types considered as {\em semantically non-essential} for elimination, specifically those with minimal impact on performance. A PDG captures the data and control dependencies between different statements of a program, providing significant semantic information. Control dependencies reflect the execution order of statements, and data dependencies reflect how variables are modified and used throughout the program. These dependencies can provide insights into the intended behavior of the program. Therefore, at the semantic level, we retained the tokens within an PDG. To assess the impact of semantic information on downstream tasks, we remove all tokens not present in the PDG of a program and measure the performance.

%This is the origin source code are listed as follows:
%\begin{lstlisting}
%public static boolean safeClose ( Closeable cb ) { 
%	if ( null != cb ) 
%		try { 
%			cb.close ( ); 
%		} catch ( IOException e ) { 
%			return false; 
%		} 
%	return true ; 
%}
%\end{lstlisting}
%Below is the result after removing statements not included in the corresponding program dependence graph.
%\begin{lstlisting}
%public static boolean safeClose ( Closeable cb ) { 
%	if ( null != cb ) 
%		try { 
%			cb.close ( ); 
%		} catch ( IOException e ) { 
%			return false; 
%		} 
%	return true ; 
%}
%\end{lstlisting}
%Note that no tokens have been removed in this example since all statements are in the PDG of this example code.

\section{Experimental Methodology}
\label{sec:methodology}

In this section, we introduce our experimental design, including downstream software engineering tasks,  models under study, datasets, evaluation metrics, and implementation.

\subsection{Downstream Tasks}
\label{sec:tasks}

We evaluated the code simplification approaches in two downstream tasks: code search and code summarization. These tasks are commonly used in software engineering to demonstrate the ability of LLMs in natural language and programming language understanding~\cite{ahmed2022multilingual,codebert-emnlp20,jiang2021treebert,wang2021syncobert,liu2022deeplearning}.
\begin{enumerate}
    \item {\bf Code Search}: This is a typical use case for pre-trained code models such as CodeBERT~\cite{codebert-emnlp20}. The goal of this task is to find relevant code snippets from a codebase given a query.
    \item {\bf Code Summarization}: The goal of this task is to generate a natural language summary for a given code snippet. It is also widely used to evaluate pre-trained models for source code~\cite{liu2019roberta}. We used this task to verify the effectiveness of program simplification in generation tasks.
\end{enumerate}

\subsection{The Models Under Study}
\label{sec:models}

%\david{here, when we talk about models, we need to talk about how codebert and codeT5 were used for code search and summarization, not just describe the in general. what the codebert-based code search appraoch look like? waht codeT5-based code search appraoch look like? what the codeT5-based code summarization look like? what the codebert-based code summarization look like?}

%\david{In one place, we may want to talk a little bit why we chose the following models. like link them with RQs and motivations.}

We opted for widely utilized models such as CodeBERT~\cite{codebert-emnlp20} and CodeT5~\cite{wang-etal-2021-codet5}, alongside the cutting-edge PLLM, GPT-4~\cite{ChatGPT}.
We aim to study the impact of code simplification on those models for code summarization and code search. Thus, the variant models under study are listed as follows:
\begin{itemize}

    \item {\em CodeBERT-based code search.}  CodeBERT is a BERT-based encoder. To perform code search, we need to add a fully connected layer on top of the CodeBERT model for binary classification.
    
    \item {\em CodeT5-based code search.} CodeT5 is a pre-trained model based on Transformer with an encoder-decoder structure. For code search, its encoder is separately extracted and joint with a fully connected layer for the classification task.

    \item {\em CodeBERT-based code summarization.}  As CodeBERT has only an encoder, it cannot perform text generation tasks. Therefore, following the work of DietCode~\cite{dietcode-fse22}, we added to it a Transformer decoder for code summarization. 

    \item {\em CodeT5-based code summarization.} As CodeT5 is suitable for text and code generation tasks, we use it directly for code summarization without any modification.

    \item {\em GPT4-based code search.} As GPT-4 provides only Web-based APIs, we cannot access its model. For classification problems like code search, we can only use prompts to get the classification results and corresponding analysis generated by GPT-4 in a specified format.

     \item {\em GPT4-based code summarization.} We input the code into the specified prompt, then submit it directly to GPT-4's Web API and receive the corresponding code description.

\end{itemize}

Currently, pre-trained large language models work under two different paradigms for natural language processing:

{\bf \circled{1} Pre-train and fine-tune paradigm}. In this paradigm, a model with a fixed architecture is first pre-trained as a Language Model (LM). Then, the pre-trained LM is adapted to different downstream tasks by introducing additional parameters and fine-tuning them using task-specific objective functions. CodeBERT and CodeT5 are the typical models used in this paradigm and also commonly used in software engineering community~\cite{codebert-emnlp20,wang-etal-2021-codet5}. 

{\bf \circled{2} Pre-train, prompt, and predict paradigm}. In this paradigm, a downstream task is reformulated to resemble those solved during the original LM training with the help of a textual prompt, rather than adapting pre-trained LMs to downstream tasks via fine-tuning. GPT-4 model is the state-of-the-art model~\cite{DBLP:journals/corr/abs-2303-08774} that can be used for code-related downstream tasks in this paradigm.

\subsection{Datasets}
\label{sec:datasets}

\begin{table}[htpb]
    \centering
    \small
    \caption{Statistics of the code snippets and code description in code search and summarization datasets (Total: the total number of code snippets, Avg/Max/Min: the average/maximum/minimal number of tokens in a code snippet or code description).}
%    \vspace{-6pt}
    \begin{tabular}{l|cccc|ccc}
        \hline
        \multirow{2}{*}{\textbf{Dataset}}& \multicolumn{4}{c|}{\textbf{Code Snippet}} & \multicolumn{3}{c}{\textbf{Code description}} \\
        \cline{2-8}
             & {\bf Total} & {\bf Avg} & {\bf Max}  & {\bf Min}  & \textbf{Avg}  & \textbf{Max}  & \textbf{Min} \\
        \hline
        Code\_Search\_{train} & 904,817 & 112.67 & 68,278 & 20 & 19.2 & 3439 & 1 \\
        \hline
        Code\_Search\_{validate} & 30,437 & 95.57 & 3,092 & 21 & 19.05 & 521 & 1 \\
        \hline
        Code\_Search\_{test} & 26,780 & 113.42 & 5,542 & 20 & 20.22 & 709& 1 \\
        \hline
        Code\_Summarization\_{train} & 164,813 & 100.99 & 512 & 17 & 13.25 & 175 & 3 \\
        \hline
        Code\_Summarization\_{validate} & 5,183 & 90.79 & 501 & 18 & 13.39 & 147 & 3 \\
        \hline
        Code\_Summarization\_{test} & 10,948 & 100.06 & 512 & 20  & 12.71 & 111 & 3 \\
        \hline
    \end{tabular}
    \label{tab:statistics_code}
%    \vspace{-5pt}
\end{table}

\subsubsection{Data Collection}

To compare results from different models in pre-train and fine-tune paradigm, we directly use the two datasets from CodeBERT~\cite{codebert-emnlp20}: code search and code summarization datasets. These datasets are the extensions of CodeSearchNet~\cite{husain2019codesearchnet}, which is a collection of datasets and benchmarks for code retrieval using natural language texts. It consists of approximately 2 millions pairs of (comment, code) that were extracted from the Github open-source repositories, covering six languages (Python, PHP, Go, Java, JavaScript, and Ruby). Our experiments were carried out only on the Java language since DietCode~\cite{dietcode-fse22} reports similar results for different languages. The statistics of the two datasets are shown in Tables~\ref{tab:statistics_code}. Each dataset was split into three portions for training, validation, and testing.

%and the datasets can be downloaded from this link\footnote{https://github.com/microsoft/CodeBERT/tree/master/CodeBERT}.

%{\bf Data Sampling.} 

\subsubsection{Data Sampling}

We have about 0.1M data points in our testing dataset. We need representative samples, as the inference of GPT-4 is expensive. Therefore, we set the confidence level of 95\%, the margin of error $E$=5\%, population proportion $p$=0.5 (as our classification task's category proportion is close to 50\%). 
The formula for computing the sample size is as follows: $n = \frac{{Z^2 \cdot p \cdot (1 - p)}}{{E^2}}$ where the $Z$ value is chosen as 1.96 because the confidence level is 95\%. That formula gives us a sample size $n$ of 384. Thus, we chose 400 as our sample size. Our sample ensures sufficient sample size as well as the time and effectiveness. The statistics of the randomly selected samples are 
shown in Table~\ref{tab:statistics_random}. As seen, the statistics of our sampled dataset are approximately consistent with that of the original datasets. For the tasks of code search and summarization, the average/max/min number of tokens in code snippets are 122.01/1680/24 and 101.27/505/22, and the average/max/min number of tokens in code description are 22.45/478/1 and 13.34/117/3.

%We randomly selected 400 samples for GPT-4 based code search and summarization models. That sample size gives the confidence level of 95\% ($Z=1.96$), $p=0.05$ and~$E=5\%$: $n = \frac{{Z^2 \cdot p \cdot (1 - p)}}{{E^2}}$ where $n=384.16 (<400)$ is the sample size. Z=1.96 is the Z value.
%p = 0.5 is the proportion of the population having the characteristics. $E=0.05$ is the margin of error. 

%\begin{comment}
\begin{table}[t]
    \centering
    \small
    \caption{The statistics of the code snippets and description of 400 random samples from code search and summarization  (Avg/Max/Min: the average/maximum/minimal number of tokens in code snippet or description.)}
    \vspace{-4pt}
    \begin{tabular}{l|c|c|c}
        \hline
        \textbf{Data} & \textbf{Avg}  & \textbf{Max}  & \textbf{Min}  \\
        \hline
        Search\_{code} & 122.01 & 1680 & 24 \\        
        \hline
        Summarization\_{code} & 101.27 & 505 & 22 \\
        \hline
        Search\_{description} & 22.45  & 478  &  1 \\
        \hline
        Summarization\_{description} & 13.34  & 117  & 3  \\
        \hline
    \end{tabular}
    \label{tab:statistics_random}
\end{table}

%\end{comment}
%The statistics of randomly sampled data are shown in Table~\ref{tab:statistics_random}.

\iffalse
From the comparison of the four tables\david{you mean table 1,2,3?}, we can see that the statistics of the random samples is basically consistent with that of the original datasets. At the same time, according to the common formula, a sample size of 400 can meet the conditions of confidence level of 98\% ($Z=1.96$), $p=0.05$ and $E=5\%$. \david{why 98\%, in SE, 95\% or 99\% is more common. or we can cite some papers to indicate that this is a commmon choice}
\[n = \frac{{Z^2 \cdot p \cdot (1 - p)}}{{E^2}}\]
where $n=384.16 (<400)$ is the sample size.
Z=1.96 is the Z value.
p = 0.5 is the proportion of the population having the characteristics.
$E=0.05$ is the margin of error.

\fi

\subsection{Metrics}
\label{sec:metrics}

The purpose of our work is to prune the input code to reduce training time while measuring the performance accordingly, given different pre-trained models and downstream tasks.
Here we use the ratio of simplification to measure the degree of simplification of a code snippet $Code$.
Given a code snippet $Code$ and its corresponding simplified one $Scode$, the simplification ratio is defined as:
\begin{equation}
    SimplifiedRatio = \frac{|Code|-|Scode|}{|Code|} \times 100%
\label{equ:ratio}
\end{equation}
where $|Code|$ and $|Scode|$ are the lengths of $Code$ and $Scode$, i.e., the number of tokens. 

The length of each code snippet and its corresponding simplified code are different. However, to have a fair comparison between different models, we need to set a fixed input size for all models. Thus, to follow the setting in DietCode~\cite{dietcode-fse22}, we set the input lengths of all original code snippets for code search and summarization to 200 and 256, respectively. Then, the input length of simplified code is $(1-SimplifiedRatio) \times 200$ or 256. 

%The efficiency of simplifying code is measured by comparing the {{\em \textbf{time cost}} it takes to achieve the same value of $r$ using paradigm 1 (fine-tuning and prediction time) and paradigm 2 (prompting and prediction time).

The efficiency of simplified code is measured by the \emph{\textbf{time cost}} it takes for fine-tuning in the pre-train and fine-tune paradigm, as well as for prediction in the prompting-prediction paradigm.

For evaluation, code search performance is measured using {\em \textbf{MRR}} (mean reciprocal rank), which is the average of the multiplicative inverse of the index for the first correct answer for the query. 
Code summarization performance is measured using the {{\em \textbf{BLEU-4}}} score, which calculates the average of $n$-gram precision on a couple of sequences. 
%with a penalty for short sequences.
%
For code search in the second paradigm (i.e., GPT-4-like models), due to the high computational requirements of MRR, we use \emph{\textbf{Precision}} which highly correlated with MRR on the models' effectiveness. Specifically, we randomly replace the code description in 400 samples and check if the replacement content matches the code part. Precision is the ratio of the number of correctly detected examples to the total number of all examples.
The ratio of matching samples to non-matching samples is 50\%.

\subsection{Implementation}
\label{sec:procedure}
We set up CodeBERT~\cite{codebert-emnlp20} and CodeT5~\cite{wang-etal-2021-codet5} with default hyper-parameters. 
For optimization, they used the Adam optimizer with learning rates of $1 \times 10^{-5}$ and $5 \times 10^{-5}$ for two downstream tasks. We used a server with 2 CPUs of Intel(R) Xeon(R) Golden 2.40GHz and 2 GPUs of Nvidia A100.

\section{Experimental results}
\label{sec:results}

\subsection{Impact of Randomly Removing Code Tokens on the Performance of PLLMs (RQ-1)}

\begin{table*}[t]
\centering
\small
% \footnotesize
\caption{Results of \underline{random-token removal} on CodeBERT and CodeT5 for \underline{code search}. (R-Time: Reduced Training Time in \%, R-MMR: Reduced MMR in \%.) (RQ1)}
%\vspace{-6pt}
\begin{tabular}{l|cc|cc|cc|cc}
\hline
%\multirow{3}{*}{\textbf{Ratio}} & \multicolumn{8}{c|}{\textbf{Code Search}} & \multicolumn{8}{c}{\textbf{Code Summarization}} \\\cline{2-17}
\multirow{2}{*}{\textbf{Ratio}}& \multicolumn{4}{c|}{\textbf{CodeBERT}} & \multicolumn{4}{c}{\textbf{CodeT5}} \\
\cline{2-9}
& \textbf{Time} & \textbf{R-Time}& \textbf{MRR} & \textbf{R-MRR}& \textbf{Time} & \textbf{R-Time}& \textbf{MRR} & \textbf{R-MRR}\\
\hline
0\%  & 433m & 0.00\%--  & 0.743  & 0.00\%-- &  434m & 0.00\%-- & 0.749 & 0.00\%--\\
10\% & 388m & 10.39\%↓  & 0.706  & 4.98\%↓  & 391m  & 9.08\%↓  & 0.737 & 1.60\% ↓\\
20\% & 354m & 18.24\%↓  & 0.694  & 6.59\%↓  & 355m  & 18.20\%↓ & 0.726 & 3.07\% ↓\\
30\% & 309m & 28.64\%↓  & 0.668  & 10.09\%↓ & 310m  & 28.57\%↓ & 0.696 & 7.08\% ↓\\
40\% & 252m & 41.80\%↓  & 0.648  & 12.79\%↓ & 253m  & 41.71\%↓ & 0.679 & 9.35\% ↓\\
50\% & 210m & 51.50\%↓  & 0.611  & 17.77\%↓ & 212m  & 51.52\%↓ & 0.641 & 14.42\% ↓\\
\hline
\end{tabular}
\label{tab:random_efficiency_code_search}
\vspace{6pt}
\end{table*}

%-------------------------------------------------------------------------------------
\begin{table*}[t]
\centering
\small
\caption{Results of \underline{random-token removal} on CodeBERT and CodeT5 for \underline{summarization}. (R-Time: Reduced Training Time in \%, R-BLEU: Reduced BLEU-4 in \%.) (RQ1)}
%\vspace{-6pt}
\begin{tabular}{l|cc|cc|cc|cc}
\hline
%\multirow{3}{*}{\textbf{Ratio}} & \multicolumn{8}{c|}{\textbf{Code Search}} & \multicolumn{8}{c}{\textbf{Code Summarization}} \\\cline{2-17}
\multirow{2}{*}{\textbf{Ratio}}& \multicolumn{4}{c|}{\textbf{CodeBERT}} & \multicolumn{4}{c}{\textbf{CodeT5}}\\\cline{2-9}
& \textbf{Time} & \textbf{R-Time}& \textbf{BLEU-4}& \textbf{R-BLEU}& \textbf{Time} & \textbf{R-Time} & \textbf{BLUE-4} & \textbf{R-BLUE}\\
\hline
0\%  & 910m& 0.00\%-- & 18.58 & 0.00\%-- & 916m & 0.00\%--& 20.49 & 0.00\%--\\
10\% & 840m& 7.69\%↓  & 17.35 & 6.62\%↓  & 845m & 7.75\%↓ & 19.56 & 4.49\%↓ \\
20\% & 802m& 11.87\%↓ & 17.18 & 7.53\%↓  & 807m & 11.90\%↓& 19.47 & 4.93\%↓ \\
30\% & 734m& 19.34\%↓ & 16.75 & 9.85\%↓  & 740m & 19.21\%↓& 19.07 & 6.88\%↓ \\
40\% & 684m& 24.84\%↓ & 16.36 & 11.95\%↓ & 689m & 24.78\%↓& 18.36 & 10.35\%↓\\
50\% & 621m& 31.76\%↓ & 15.06 & 18.95\%↓ & 627m & 31.55\%↓& 17.30 & 15.53\%↓\\
\hline
\end{tabular}
\label{tab:random_efficiency_summarization}
\vspace{6pt}
\end{table*}
%----------------------------------------------------------------------------------

%{\bf Analysis Procedure.}

%\david{in the section of analysis procedure, please add how exactly you did the experiments in this rq}

\subsubsection{Analysis Procedure}

To answer this research question, we randomly removed code tokens from each given input code snippet, and then the simplified code is fed into CodeBERT and CodeT5 for evaluating the performance metrics, i.e., MRR and BLEU-4, and the corresponding training time. Particularly, for each code snippet in training, validation and testing datasets, we first calculated the length of its tokens. Then, we randomly removed 10\%-50\% of the total tokens and fed the simplified code input into the CodeBERT and CodeT5 models.

% with the input length that calculated by Equation~\ref{equ:ratio}.
%To answer \underline{RQ-1}, we randomly remove code tokens from the code snippets with various simplification ratio $r$ to check the changes of training time, MRR and BLEU-4 of CodeBERT and CodeT5 for code search and summarization tasks respectively. 
%\david{how exactly you did this? In training, how did you remove? in prediction, how did you remove? discuss the detailed analysis proceduce in detail, and blalallalla}

%we use random tokens with various simplification ratio $r$ to check the changes of training time, prediction time, MRR and BLEU-4 of CodeBERT and CodeT5 for code search and summarization tasks respectively. 

%{\bf RQ1-Results.} 

\subsubsection{Empirical Results}

Tables~\ref{tab:random_efficiency_code_search} and~\ref{tab:random_efficiency_summarization} show the empirical results on the positive correlations between the simplified ratio and training time, and those between the change of MRR and BLUE-4 of both CodeBERT and CodeT5 models. 
For instance, {\em when 50\% of the code is removed, the training time for code search using the CodeBERT model decreases by 51.50\%}. Similarly, for code summarization using CodeT5, {\em the training time is reduced by 31.55\% when the simplification ratio is set to 50\%}.

%Tables~\ref{tab:random_efficiency_1} shows 1) the relationship between the simplified ratio and the training time of CodeBERT and CodeT5 models respectively, and 2) the change of MRR and BLUE-4 of both models. 

%From the three tables, as the simplification ratio increases,
%\begin{enumerate}
    %\item The training time follows a linear-like descent. 
    %\item The performance in terms of MRR and BLUE-4 of both CodeBERT and CodeT5 appropriately decreases and CodeT5 outperforms CodeBERT.
%\end{enumerate}
%The visualization of the above Table~~\ref{tab:random_efficiency_1} and~~\ref{tab:random_effectiveness_1} are shown in Figure~\ref{fig:random_overview}.

\subsubsection{The Impact of Random-Token Removal on CodeBert and CodeT5}

%{\bf The impact of random-token removal on CodeBert and CodeT5.} 

For CodeBERT, as the simplified ratio increases, {\em the training time follows a linear-like descent} on both of downstream tasks. When 10\%-50\% of the code tokens is removed, the training time of CodeBERT on code search can be reduced by 10.39\%-51.5\% with a sacrifice of MRR in dropping 4.98\%-17.77\%. The percentage of reduced time is always greater than the performance sacrifice for CodeBERT on code search. {\em When the code simplification ratio increases (i.e., more code tokens are removed), the percentage of the training time saved also increases, but the speed of performance declines slower than the increase speed of the code simplification ratio}. We observe the similar trend for CodeBERT on code summarization.
This could serve as a guidance for practitioners in the code simplification process.

%However, the decline in code search is much faster than that in code summarization. For example, 
%{\color{brown}{For CodeBERT, as the simplification ratio increases, the training time follows a linear-like descent on both of downstream tasks. however, the decline in code search is much faster than in code summarization.  }}

%{\bf The impact of random code token removal on CodeT5.} {\color{brown}{CodeT5 achieves similar time-saving results as CodeBERT, but with better MRR and BLUE-4 performance in the same conditions. }}

%{\bf CodeBERT vs. CodeT5.} 

\subsubsection{CodeBERT versus CodeT5}

CodeT5 achieves similar time-saving results as CodeBERT, but with better MRR and BLUE-4 performance in the same conditions. We observe the differences in the results from both models on two tasks. CodeBERT has only one encoder, while CodeT5 has an encoder-decoder structure. 
The pre-training of CodeT5 is much more complex than CodeBERT, and it could learn more correspondences between tokens of code snippets and the descriptions. Therefore, when removing random tokens, the CodeT5 model can also find more tokens with "hints"  in the remaining code to help improve the effectiveness of downstream tasks.

For code search, the encoders of CodeBERT and CodeT5, and one fully-connected layer, are used for matching checks. The time complexity of the core module is $4LC^2 + 2L^2C$, where $L$ is the length of the input code and $C$ is the size of the dimension of each token. In our case, $C=768$, and the average $L$ is less than 120 (Table~\ref{tab:statistics_code}). Thus, the first factor of time complexity is much larger than the second one. When $L$ decreases, the change in time follows a pattern similar to a linear curve. %as shown in Figure~\ref{fig:random_overview}.

For code summarization, CodeBERT has added text generation capability with a Transformer decoder as in DietCode~\cite{dietcode-fse22}. As the decoder also requires a lot of training time, this makes the total training time for code summarization longer. Moreover, the proportion of time saved by simplifying the code can be lower in the total training time. This is consistent with the result in Table~\ref{tab:random_efficiency_code_search}.

%\begin{boxB}
%    Here we have the following findings: Reducing the amount of code input can effectively shorten the running time. On code search, the ratio of time reduction to code reduction is almost the same, while on code summarization, the ratio of time reduction to code reduction is about 0.6. At the same time, the decrease in model performance is also reasonable.
%\end{boxB}

Moreover, from the columns {\em Ratio} and {\em R-Time} in Table~\ref{tab:random_efficiency_code_search}: (10\%,10.39\%), (20\%,18.24\%), (30\%,28.64\%), etc., the ratio is 1:1 between time saved and code reduction ratio for code search. Similarly, the corresponding ratio is 1:0.6 for code summarization as observing the same columns in Table~\ref{tab:random_efficiency_summarization}. 

\intuition{
{\bf RQ1 Takeaway}: 
Minimizing the volume of code entered can significantly cut down on training time. In the context of code search, the correlation between time saved and code reduction is nearly one-to-one (1:1). However, in the realm of code summarization, this correlation is approximately 0.6 (0.6:1), indicating a slightly lower reduction in time compared to code reduction.
%Reducing the amount of code input can effectively save the training time. On code search, the ratio of time reduction to code reduction is almost the same (i.e., 1:1), while on code summarization, the ratio of time reduction to code reduction is about 0.6 (i.e., 0.6 : 1). 
With the same reduction in the input, the performance of CodeT5 is better than that of CodeBERT.
}

\subsection{Impact of Removing Lexical Tokens on the Performance of PLLMs (RQ-2)}

\subsubsection{Analysis Procedure}

We removed all of the identifiers and symbol tokens from each code snippet in the training, validation, and testing datasets for CodeBERT and CodeT5. We then checked their training time and performance on both downstream tasks. Particularly, we used JavaParser~\cite{ast-maven} to build AST, then identified the locations of the identifiers in the code based on the \code{NameExpr} node in the AST. For simple symbols, we deleted them from the code snippets using string matching.

%{\color{brown}{For each code snippet in the training, validation, and testing datasets, we remove identifiers and symbol tokens completely. We then check their training time and performance on both downstream tasks using CodeBERT and CodeT5 models. More specific, for identifiers, we use the JavaParser tool~\cite{ast-maven} to convert each code snippet into an AST, then identify the locations of the identifiers in the code snippet based on the NameExpr node in the AST and finally remove it. For simple symbols, we simply delete them from the code snippets.}}

%{\bf RQ2-Results.} 

\subsubsection{Empirical Results}

Tables~\ref{tab:categorized-code-search} and ~\ref{tab:categorized-code-summarization} display the results on the impact of removing lexical tokens including {\em identifiers and symbols} on code search and code summarization. For instance, removing all identifiers (accounting for 15.48\% of the code tokens) leads to a 12.52\% decrease in MRR for code search using the CodeBERT model. On the contrary, for symbol tokens, increasing the simplified ratio to 52.65\% in the code summarization task led to only 0.59\% decrease in BLEU-4.

%{\color{brown}{Table~\ref{tab:categorized} presents the effects of removing lexical tokens on code search and code summarization, respectively. For instance, removing all identifiers (account for 15.01\%) resulted in a 12.52\% decrease in MRR for code search using the CodeBERT model. On the contrary, for symbol tokens, increasing the simplified ratio to 52.65\% in the code summarization task only led to a 0.59\% decrease in BLEU-4.}}

\subsubsection{Impact of Identifier Removal}

The number of identifiers accounts for 15.48\% of total tokens. Removing all identifiers can save training time about 15\% and 9\% for both models on code search and summarization, respectively, but with a sacrifice of performance. The performance sacrifice of a model on two tasks are different: CodeBERT drops 12.52\% on code search and 3.83\% on code summarization; CodeT5 drops 8.81\% on code search and 4.88\% on code summarization, indicating that {\em identifiers can affect code search more than code summarization}.

%the code search task is more sentivive to identifiers are more useful

%-----------------------------------------------------------------------------------
\begin{table*}[t]
\centering
\small
\caption{Impact of categorized tokens on \underline{code search} for CodeBERT and CodeT5, measured by training time and MRR. (R-Time: Reduced Training Time in \%, R-MMR: Reduced MMR in \%.) (RQ2)}
\vspace{-5pt}
\resizebox{\linewidth}{!}{
\begin{tabular}{l|l|cc|cc|cc|cc}
\hline
%\multirow{3}{*}{\textbf{Methods}}&  \multicolumn{9}{c}{\textbf{Code Search}} \\\cline{2-10}

Methods & \multirow{2}{*}{\textbf{Ratio}} &\multicolumn{4}{c|}{\textbf{CodeBERT}} & \multicolumn{4}{c}{\textbf{CodeT5}} \\\cline{3-10}
 & &\textbf{Time} & \textbf{R-Time}& \textbf{MRR} & \textbf{R-MRR}& \textbf{Time} & \textbf{R-Time}& \textbf{MRR} & \textbf{R-MRR}\\
\hline
Base              & 0.00\%   & 433m & 0.00\%-- & 0.743 & 0.00\%-- &  434m & 0.00\%-- & 0.749 & 0.00\%--\\
Identifiers       & 15.48\%  & 367m & 15.24\%↓ & 0.650 & 12.52\%↓ & 375m  & 13.59\%↓ & 0.683 & 8.81\%↓\\
Symbol tokens     & 51.38\%  & 215m & 50.35\%↓ & 0.722 & 2.83\%↓  & 193m  & 55.53\%↓ & 0.729 & 2.67\%↓\\
Control structures & 11.90\%  & 381m & 12.01\%↓ & 0.715 & 3.77\%↓  & 386m  & 11.06\%↓ & 0.730 & 2.54\%↓\\
Method invocations & 37.47\%  & 266m & 38.57\%↓ & 0.682 & 8.21\%↓  & 268m  & 38.25\%↓ & 0.698 & 6.81\%↓\\
Method signature  & 15.96\%  & 354m & 18.24\%↓ & 0.649 & 12.65\%↓ & 354m  & 18.43\%↓ & 0.680 & 9.21\%↓\\
PDG               & 24.84\%  & 332m & 23.33\%↓ & 0.713 & 4.04\%↓  & 328m  & 24.42\%↓ & 0.729 & 2.67\%↓\\
\hline
\end{tabular}
}
\label{tab:categorized-code-search}
%\vspace{1pt}
\end{table*}
%-----------------------------------------------------------------------------------
\begin{table*}[t]
\centering
\small
\tabcolsep 3pt
\caption{Impact of categorized tokens on \underline{code summarization} for CodeBERT and CodeT5, measured by training time and BLEU-4. (R-Time: Reduced Training Time in \%, R-BLEU: Reduced BLEU-4 in \%) (RQ2)}
\vspace{-8pt}
%\resizebox{\linewidth}{!}{
\begin{tabular}{l|l|cc|cc|cc|cc}
\hline
%\multirow{3}{*}{\textbf{Methods}}&  \multicolumn{9}{c|}{\textbf{Code Search}} & \multicolumn{9}{c}{\textbf{Code Summarization}} \\\cline{2-19}
Methods & \multirow{2}{*}{\textbf{Ratio}} & \multicolumn{4}{c|}{\textbf{CodeBERT}} & \multicolumn{4}{c}{\textbf{CodeT5}}\\\cline{3-10}
 & &\textbf{Time} & \textbf{R-Time}& \textbf{BLEU-4}& \textbf{R-BLEU}& \textbf{Time} & \textbf{R-Time} & \textbf{BLUE-4} & \textbf{R-BLEU}\\
\hline
Base              & 0.00\%   & 910m      & 0.00\%--  & 18.58 & 0.00\%-- & 916m & 0.00\%--& 20.49& 0.00\%--\\
Identifiers       & 15.69\%  & 829m      & 8.90\%↓& 17.87& 3.83\% ↓& 828m & 9.61\% ↓& 19.49& 4.88\% ↓\\
Symbol tokens     & 52.31\%  & 606m      & 33.41\%↓& 18.47& 0.59\% ↓& 628m& 31.44\% ↓& 20.34& 0.73\% ↓\\
Control structures & 16.48\%  & 820m      & 9.89\%↓& 18.57& 0.05\% ↓& 843m & 7.97\% ↓& 20.38& 0.54\% ↓\\
Method invocations & 35.48\%  & 692m      & 23.96\%↓& 18.17& 2.21\% ↓& 703m& 23.25\% ↓& 20.31& 0.88\% ↓\\
Method signature  & 11.36\%  & 813m      & 10.66\%↓& 15.86& 14.64\% ↓& 817m& 10.81\% ↓& 16.61& 18.94\% ↓\\
PDG               & 23.62\%  & 750m      & 17.58\%↓  & 18.46& 0.65\% ↓& 766m& 16.38\% ↓& 20.39& 0.49\% ↓\\
\hline
\end{tabular}
%}
\label{tab:categorized-code-summarization}
%\vspace{-11pt}
\end{table*}

%The number of identifiers is second to last, accounting for only about 15\%. This 15\% proportion of identifiers caused CodeBERT and CodeT5 to drop by -12.52\%, -8.81\%, -3.84\%, and -4.83\% in code search and code summarization respectively.

\subsubsection{Impact of Symbol Token Removal}

The number of symbols accounts for +50\% of the total tokens. Removing all the symbols can save the training time by 51\% and 31\% in both models (based on the RQ1 results) on code search and summarization, respectively, but with a performance decrease. The performance sacrifice on two tasks are consistent: CodeBERT only drops 2.83\% on code search and 0.59\% on summarization; CodeT5 only drops 2.67\% on code search and 0.73\% on summarization, indicating that {\em symbols have less impact on code search and code summarization}.

%{\color{brown}{Symbol tokens make up the largest proportion of all types of tokens in code snippets, accounting for almost 50\%. However, their corresponding impact is much smaller than that of identifiers, with the corresponding drops -2.83\%, -2.67\%, -0.05, and -0.49\%.}}

\subsubsection{Symbols versus Identifiers}

The notable difference in the impacts of symbols and identifiers could be attributed to the following factors: 1) Identifiers exhibit a closer resemblance to natural language texts compared to symbols. This proximity to the source code's description imparts greater cues for both classification and generation tasks. 2) Typically, identifiers are closely linked to a method's name, which encapsulates its purpose or functionality~\cite{icse20}. Consequently, identifiers display a clear correlation with certain tokens in the code description. This correlation leads to numerous many-to-many correspondences following the fine-tuning process, thereby enabling identifiers to furnish a substantial array of indications for subsequent tasks.

\begin{comment}
From the tables, we have the following observations:
\begin{enumerate}
    \item Symbol tokens have the largest proportion among all types of tokens in code snippets, accounting for almost 50\%. Meanwhile, the number of identifiers is second to last, accounting for only about 15\%. 
    \item The 15\% proportion of identifiers caused CodeBERT and CodeT5 to drop by -12.52\%, -8.81\%, -3.84\%, and -4.83\% in code search and code summarization respectively. 
    \item The corresponding effect of symbol tokens, which account for 50\% of the tokens, is only -2.83\%, -2.67\%, -0.05, and -0.49\%, which is much smaller than the effect of identifiers.
    %\item In the same case, the performance drop of CodeBERT is higher than that of CodeT5.
\end{enumerate}
\end{comment}

%CodeT5's pre-training tasks are much more comprehensive and complex than CodeBERT's, so CodeT5 can learn more correspondence between tokens, resulting in smaller performance degradation.

%In the study of Dietcode~\cite{}, the authors argue that separators and mathematic operators obtain more weights from the encoder attention and thus they should be given priority to be reserved. However, this  This does not match our experimental results. In fact, the token weights of the encoder cannot fully represent the importance of tokens for downstream tasks. We will discuss this in future work.

\intuition{
{\bf RQ2 Takeaway}: 
Making up a substantial portion of the content yet have minimal impact on performance, symbols are well-suited for code removal. Conversely, while identifiers constitute a smaller proportion, they wield a more significant influence on the model's performance. Therefore, a decision to retain or eliminate identifiers should be made at a later stage in the code simplification.
%Due to the large proportion of symbol tokens and their small impact on the model's performance, they are suitable for removal from the code. On the other hand, identifiers have a small proportion but a greater impact on the model's performance, so they should be kept or removed later in the process of code simplification.
}

\subsection{Impact of Removing Syntactic Tokens on the Performance of PLLMs (RQ3)}

\subsubsection{Analysis Procedure}

We removed the tokens from the following syntactical structures: {\em control structures, method invocations}, and {\em method signatures}. For control structures, we used the JavaParser tool to convert each code snippet into an AST. Then, we identified the \code{for}, \code{if}, \code{try}, \code{catch}, \code{switch}, \code{while}, and \code{do-while} statements in the code segment by extracting the \code{ForStmt}, \code{IfStmt}, \code{TryStmt}, \code{CatchStmt}, \code{SwitchStmt}, \code{WhileStmt}, and \code{DoStmt} nodes from the AST. We then recognized their condition expressions based on parentheses, and finally removed all of them. The handling of method invocations and signatures is similar to that of control structures.

\subsubsection{Empirical Results}

Tables~\ref{tab:categorized-code-search} and ~\ref{tab:categorized-code-summarization} show the impact of removing syntactic tokens on code search and code summarization. For instance, with a simplified ratio of 18.05\% for CodeT5 on code summarization, there is an 11.36\% drop in BLUE-4 score for token removal in method signatures. Similarly, for CodeBERT and the removal of tokens in method invocations and control structures, with a simplified ratio of 37.47\%/11.90\%, there is an 8.21\%/3.77\% drop in MRR in code search.

%{\color{brown}{Table~\ref{tab:categorized} shows the impacts of removing syntactic tokens on code search and code summarization respectively. For instance, with a simplified ratio of 18.05\% for CodeT5 on the code summarization, there is an 18.94\% drop in BLUE-4 score for tokens in method signature. Similarly, for tokens in method invocation/control structure, with a simplified ratio of 38.61\%/11.76\%, there is an 8.21\%/3.77\% drop in MRR.}}

\subsubsection{Impact of the Removal of the Tokens in Method Signatures} 

For both models, tokens in method signatures have the greatest impact on code search and summarization. For example, the removal of the signature tokens that account for 11.36\% of all tokens in the code summarization dataset, reduces the BLEU score by 14.64\% and 18.94\% for CodeBERT and CodeT5, respectively. Additionally, their impact on code search is very similar to that of identifiers, while their impact on code summarization is significantly stronger compared to all other tokens, regardless of the model.

%{\bf The impact of the removal of the tokens in method signatures.} For both models, tokens in method signatures have the greatest impact on code search and summarization.(i.e, the signature tokens, which account for 15.96\%/11.36\% of all tokens, can reduce MRR/BLEU-4 by 12.65\%/14.64\% and 9.21\%/18.94\% for CodeBERT and CodeT5 respectively in the code search/summarization dataset), even more than identifiers(i.e., the identifiers, which account for 15.48\%/15.69\% of all tokens, can reduce MRR/BLEU-4 by 12.52\%/13.83\% and 8.81\%/4.88\% for CodeBERT and CodeT5 respectively in the code search/summarization dataset). Additionally, their impact on code search is very similar to that of identifiers, while their impact on code summarization is significantly stronger compared to all other tokens, regardless of the model. 
%\david{no need to rewrite here, I need concrete numbers to support the argument, just one line or one result.}

%{\bf The impact of the removal of the tokens in method signatures.} {\em Tokens in method signatures have the greatest impact on code search and for both models, even surpassing identifiers. Meanwhile, their impact on code search is very close to that of identifiers, regardless of the model.}

\subsubsection{Impact of the Removal of the Tokens in Method Invocations and Control Structures} 

The impacts of tokens in method invocations and control structure are similar. For example, the tokens in method invocations and control structures account for 37.47\% and 11.90\% of all tokens in the code search dataset, and the removal of them reduces MRR by 8.21\% and 3.77\% for CodeBERT, respectively. Thus, the tokens in control structures have slightly more impact than tokens in method invocations. Removing tokens in control structures and method invocations has a higher impact on code search compared to that of code summarization.
%the tokens in control structures 
%can reduce MRR by 3.77\% and 2.54\% for CodeBERT and CodeT5, respectively.

%\david{same here, just add some summarization of stats here, no need to rewrite here. It is too dry if only text without numbers}

%Additionally, tokens have a greater impact than symbol tokens, but a lesser impact than tokens in method signature.

\subsubsection{Difference of Impacts among Tokens in Control Structures, Method Signatures and Invocations}

The most profound impact is attributed to the tokens within method signatures. For example, the removal of the signature tokens that account for 11.35\% of the whole code can cause a performance drop of 18.94\% in BLEU score for CodeT5 on summarization. In typical instances, the names and formal parameters of regular methods are meticulously crafted, encompassing information that surpasses mere syntax, encapsulating the method's intention and operations. This fixed co-occurrence relationship between method signatures and their corresponding descriptive tokens facilitates the assimilation of their correlation by models. Consequently, tokens present in method signatures offer richer cues for downstream tasks, notably in steering the process of description generation. 
Tokens within method invocations and control structures hold significance due to their strategic positions within the AST, rendering the identification of corresponding tokens within the description a more straightforward endeavor. Notably, crucial branch conditions tend to be mirrored by pertinent tokens within the description. %\david{this part is too dry, do we have results or data to support the above paragraph? it would be convinving to give some stats.}

%The tokens in method signatures have the greatest impact. The names and formal parameters of regular methods are usually carefully designed, containing information that goes beyond syntax, such as the method's purpose and functionality. Because there is a fixed co-occurrence relationship between the method signature and the corresponding description tokens, the correspondence between them can be learned by models. As a result, tokens in the method signatures provide more hints for downstream tasks, particularly in guiding description generation. For tokens in method invocations and control structures, their important positions in the AST make them easier to find corresponding tokens in the description. For example, important branch conditions are likely to have corresponding tokens in the description.

\intuition{
{\bf RQ3 Takeaway}: 
The tokens within method signatures wield the most substantial influence over downstream tasks, particularly those pertaining to generation tasks. Tokens in control structures and method invocations possess a moderate impact on downstream tasks, surpassing symbol tokens by a significant margin and even outweighing the influence of identifiers. Tokens in method invocations
and control structures have a higher impact on code search than on code summarization.
%Finally,they should be removed after symbol tokens are removed, and AST does help to find tokens with 'hints'.
}

%\subsection{(RQ-4) Impact of removing semantic tokens on the performance of PLLMs}

\subsection{Impact of Removing Semantically Non-Essential Tokens (RQ4)}

%{\bf Analysis Procedure.}

\subsubsection{Analysis Procedure}

An PDG captures the interconnections of data and control among various segments of source code, offering substantial semantic insights. These inter-dependencies can unveil the envisaged program behavior. The PDG serves as a vehicle for representing source code semantics, with lesser significance placed on statements lying outside its scope in the code's representation.
We use JAST2DyPDG~\cite{pdg-tool} to convert each code snippet into a program dependence graph. Unlike the experiments in \underline{RQ-(2-3)}, we do not remove the tokens from a PDG. We remove the nodes (statements) that are not involved in either of two relations: data and control dependencies. We evaluate the training time and the performance of CodeBERT and CodeT5 on the simplified~code. 

%{\bf RQ4-Results.}

\subsubsection{Empirical Results}

Tables~\ref{tab:categorized-code-search} and ~\ref{tab:categorized-code-summarization} show the impact of removing tokens not in PDG on code search and summarization. For example, removing all tokens not in PDG (account for 24.84\% of total tokens) leads to a 4.04\% decrease in MRR for code search using CodeBERT. Meanwhile, the simplified ratio of  23.62\% in the code summarization only leads to a 0.49\% decrease in BLEU-4 for CodeT5, which indicates that the tokens outside the PDG is not important. We further analyzed the tokens outside the PDG and observed that most of them are in the code structures such as \code{try}, \code{catch}, and \code{finally}.

%The main reason of the tokens outside of PDG having less impact is that they lack the main information of the method, many code structure tokens outside the PDG, such as \code{try}, \code{catch}, and \code{finally}.
%These make it difficult to find corresponding tokens in the description.

\begin{comment}
    From the table, we can see that 
\begin{enumerate}
    \item In both datasets, the proportion of tokens outside the PDG is relatively moderate, and many code structure tokens outside the PDG, such as \code{try}, \code{catch}, and \code{finally}.
    \item The impact of tokens outside PDG is less significant than that of tokens in control structure and invocation, but more significant than that of symbol tokens for both the CodeBERT and CodeT5 models on code search and summarization, respectively. 
\end{enumerate}

The main reason why tokens outside of PDG have little impact is that they lack the main information of the method, making it difficult to find corresponding tokens in the description.
\end{comment}

\intuition{
{\bf RQ4 Takeaway}: 
Tokens that are not present in PDGs have little impact on downstream tasks due to their little semantic information. The impact of tokens outside the PDG is less significant than that of tokens in control structures and invocations, but more significant than that of symbol tokens for both the CodeBERT and CodeT5 models on code search and code summarization. 
}

%\subsection{(RQ-5) Can we propose an effective code simplification method?}

\section{{\tool}:  Model-Agnostic Code Simplification for PLLMs}
\label{sec:algorithm}

%\textbf{Analysis Procedure.} 
%To answer (\underline{RQ-5}), we must what we have learned from \underline{RQ-(2-4)}.

%Based on the The main objective of \underline{RQ-(2-4)} is to recognize tokens, i.e., a token should be retained as much as possible if it provides more ‘hints’ for solving downstream tasks. Otherwise, it should be removed as earlier as possible.
%Based on the presented experimental results, we can establish a basic ranking policy for tokens. It can prioritize each token in a code snippet according to its category in the following order: \textbf{method signature > identifiers > control structure = method invocation > symbol tokens}. 

\subsection {Important Levels of Tokens} 

The crux of what we have learned from the previous experiments for RQ-(2-4) is the relative order of the importance of 
each type of tokens on the performance of the models in the downstream tasks. We have the order of importance as follows: \textbf{method signature > identifiers > control structures $\approx$ method invocations > symbol tokens}. This relative order is model-agnostic and even task-independent. Note that 1) although the tokens in control structures are slightly better than the ones in method invocation, we still consider them almost equal; 2) tokens not in PDG are not included in this ranking since building PDGs is time-consuming and might not always possible for incomplete code.

\begin{table}[h]
% \footnotesize
\centering
%\scriptsize
\small
\caption{The overlap of the categories of tokens ( check/cross/line mark means that there is/isn't an overlap/meaningless in the two corresponding categories)}
\vspace{-6pt}
%\resizebox{\linewidth}{!}{
\begin{tabular}{l|c|c|c|c|c}
    \hline
                & Signature     & Identifiers    & Invocations          & Structures            & Symbols \\
    \hline
    Signature   & --            & --            & --                  & --                   & --  \\
    \hline
    Identifiers  & $\checkmark$  & --            & --                  & --                   & -- \\
    \hline
    Invocations  & $\times$      & $\checkmark$  & --                  & --                   & -- \\
    \hline
    Structures   & $\times$      & $\checkmark$ & $\checkmark$        & --                   & -- \\
    \hline
    Symbols      & $\checkmark$  & $\times$      & $\checkmark$        & $\checkmark$         & --  \\
    \hline
\end{tabular}
%}
% \vspace{5pt}
\label{tab:overlap}
\vspace{-5pt}
\end{table}

The principle of our novel model-agnostic code simplification technique is that {\em the tokens with the lower levels of importance
should be removed before the ones with the higher levels}. The ranking score for the important level of a token is based on
its category. An out-of-category token is the least important one. A token can belongs to multiple categories (Table~\ref{tab:overlap}). 
In some cases, a token may belong to more than one categories, which actually are more important than the ones belonging to one category. For example, an identifier can appear in a method invocation or a control structure. 
%If an identifier belongs to one of these two cases, then its ranking score is the sum of its categories' ranking scores since it becomes more important in such cases.

Based on the above principle, we assign different rank scores to various tokens. Tokens with a lower rank score have higher importance. For example, the rank score of "signature" is the highest at 1, followed by identifiers in a control structure or method invocation with scores of 2 and 3. 
Identifiers in signature receive a ranking score of 1, as they are integral parts of the signature. Symbol tokens within the signature receive a ranking score of 7, emphasizing their uniform importance across all instances. Tokens in control structures and  method invocations are not shown in Table~\ref{tab:overlap}. Their ranking scores are assigned with 6 and 5, respectively since control structures slightly outweigh method invocations. Other tokens are assigned with the score of 8.

%The rank scores of identifiers, tokens in control structures and method invocations, symbol tokens, and others are 4, 5, 6, 7, and 8, respectively.

%With the basic rank, we can set a ranking score for each category. Then the ranking score of each token is based on its category. 
%With the ranking score, we can do code simplification in greedy manner, i.e., tokens with higher rank scores should be removed later than the ones in lower rank score.(An out-of-category token has the lowest ranking score here.)
%However, a token can belongs to multiple categories, see Table~\ref{tab:overlap}).
%In some cases, tokens that belong to two categories may be more important than tokens that belong to only one type.
%Here we focus on the cases where identifiers appear in method invocation, method signature, and control structure. 
%If an identifier belongs to one of these three cases, then its ranking score is the sum of its categories' ranking scores since identifiers become more important in such cases.

\subsection{Problem Formulation for Code Simplification}

\begin{algorithm}[t]
    \caption{{\tool}: Code Simplification Algorithm.}
    \label{alg:slimcode}
\begin{flushleft}
        \textbf{INPUT:} $D = \{d_1,...,d_m\}$, ranking scores $V$, $SimplifiedRatio$ and the original input length $L$ \\
        \textbf{OUTPUT:} A simplified code dataset $D'$\\
        \textbf{PROCEDURE:} 
\end{flushleft}
    \begin{algorithmic}[1]
    \State Initialize $D'$, a copy of $D$
    \For {$j$ from 1 to $m$}
        %\State Initialize a list $d'_j= \{t_1,...,t_{n_{j}} \}$ copying $d_j \in D$ 
        \State Initialize an empty dictionary $removedTokens$ with positions and their tokens
        \If{$n_j > (1-SimplifiedRatio) \times L$}
            \State $\mathcal{W} \gets n_j - $(1-SimplifiedRatio)$ \times L$
            \State $currentWeight \gets 0$
            \While {$currentWeight < \mathcal{W}$}
                \State Add \{index: token with highest $v$ \} ($\in$ $d'_j$,$\notin$ $removedTokens$) into $removedTokens$
                \State $currentWeight \gets sizeof(removedTokens)$
            \EndWhile
            \State $d_j = d_j/selectedTokens[1:\mathcal{W}]$
        \EndIf
    \EndFor
    \State \textbf{return} $D'$
    \end{algorithmic}
\end{algorithm}

A dataset contains $m$ snippets (documents) and each code snippet $d_j$ ($ 1\leq j \leq m$) can be considered as a sequence of tokens, denoted as $d_j = \{t_1,...,t_{n_{j}}\}$. The index $i$ of $t_i$ records the position of the corresponding token in the code snippet.
The same token can appear multiple times, i.e., $\exists i, j$ ($i \neq j$) and $t_i=t_j$.
Each token $t_i$ ($ 1\leq i \leq n_j$) represents an item. 
$v_i$ represents the ranking score of each token $t_i$, and the weights $w_i$s of all tokens in $d_j$ are equal to 1. 
With the given $SimplifiedRatio$, the total number of tokens to be removed for each snippet is $\mathcal{W} = n_j-(1-SimplifiedRatio) \times L$ and $L$ is the original input length.

%Strictly speaking, firstly, it is better to model a code snippet as a sequence of tokens rather than a set of tokens, denoted as $d_j = \{t_1,...,t_{n_{j}}\}$. The index $i$ of $t_i$ records the position of the corresponding token in the code snippet.
%Secondly, the same token can appear multiple times, i.e., $\exists i, j$ ($i \neq j$) and $t_i=t_j$.

%We remove the tokens in one code snippet at a time.
%Therefore, 

%Despite these issues, our case still satisfy the definition of Equation~\ref{def:kp}.

We formulate code simplification as an 0-1 knapsack optimization problem. Given a set of $n$ items numbered from 1 up to $n$, each with a weight $w_i$ and a value $v_i$, along with a maximum weight $\mathcal{W}$, the knapsack problem is defined in Equation~\ref{def:kp} as follows:
\begin{equation}
\begin{split}
& maximize \sum_{i=1}^{n} v_ix_i, \ such \  that \sum_{i=1}^{n} w_ix_i \leq \mathcal{W} \  and \  x_i \in \{0,1\}.
%& such \  that \sum_{i=1}^{n} w_ix_i \leq \mathcal{W} \quad and \quad x_i \in \{0,1\}.
\end{split} 
\label{def:kp}
%\vspace{-5pt}
\end{equation}
\subsection{{\tool} Algorithm}

Algorithm~\ref{alg:slimcode} shows our greedy method ({\tool}) for code simplification.
In Algorithm~\ref{alg:slimcode}, at lines 1, {\tool} initializes a copy of the original code dataset $D'$ as the returned simplified code dataset. At lines 2-10, it removes the tokens of each snippet in $D'$ one-by-one based on their ranking scores stored in the dictionary $V$. 
In line 3, \code{removedToken} records the pair of the index and the corresponding tokens with the highest rank score (\{index:token\}) in $d'$ at each turn.
In line 4, $(1-SimplifiedRatio) \times L$ is the number of tokens to be retained.
Line 5 defines the number of removed tokens.
In line 6, \code{currentWeight} represents the number of currently selected tokens to be removed. In lines 7-9, at each turn, the algorithm repeatedly selects the remaining token (not in \code{removedToken}) with the highest ranking score from left to right until the number of removed tokens is reached.
Finally, each code removes the first $\mathcal{W}$ selected tokens (line 10).

\subsection{Empirical Evaluation Results on {\tool}}
\label{slimcoderesults}

%Table~\ref{tab:SlimCodeResult} and Table~\ref{tab:DietCodeResult} show the results of {\tool} and DietCode in code search and code summarization. The tables demonstrate that {\tool} outperforms the state-of-the-art DietCode~\cite{dietcode-fse22} in all experiments.

\begin{table}[t]
\setlength{\tabcolsep}{3pt}
\centering
\small
\caption{Results of \underline{{\tool}} for CodeBERT and CodeT5 on code search and summarization.(10\%-50\%:removing 10\%-50\% tokens for each snippet. Time: time to running removal algorithms. Times:the multiple by which {\tool} is faster than DietCode.) }%(DietCode provides only the result at a rate of 40\%.)}
\vspace{-4pt}
\begin{tabular}{l|cc|cc|cc|cc|cc|cc}
\hline
 \multirow{3}{*}{\textbf{Ratio}}& \multicolumn{6}{c|}{\textbf{Code Search}} & \multicolumn{6}{c}{\textbf{Code Summarization}} \\
\hline
 & \multicolumn{2}{c|}{\textbf{CodeBERT}} & \multicolumn{2}{c|}{\textbf{CodeT5}} & \multicolumn{2}{c|}{\textbf{Pruning}} & \multicolumn{2}{c|}{\textbf{CodeBERT}} & \multicolumn{2}{c|}{\textbf{CodeT5}} & \multicolumn{2}{c}{\textbf{Pruning}}\\
 & MRR & R-M &  MRR & R-M & Time & Times & BLUE & R-B  & BLUE & R-B & Time & Times\\
\hline
Base &  0.743 &  0.00\% -- &0.754 &0.00\%--   &N/A  &------   &18.58 &0.00\%-- &20.49 &0.00\%-- &N/A &------    \\
10\% &  0.740 &  0.82\% ↑  &0.749 &0.66\%↓    &17m  &32.2    &18.56 &0.11\%↓  &20.44 &0.24\%↓  &45s  &133.1     \\
20\% &  0.721 &  1.77\% ↓  &0.748 &0.80\%↓    &17m  &28.9    &18.29 &1.56\%↓  &20.38 &0.54\%↓  &53s  &101.4     \\
30\% &  0.734 &  0.00\% --  &0.751 &0.40\%↓    &20m  &21.9    &18.62 &0.22\%↑  &20.49 &0.00\%↓  &59s  &80.3      \\
40\% &  0.733 &  0.14\% ↓  &0.757 &0.40\%↑    &21m  &18.3    &18.41 &0.91\%↓  &20.52 &0.15\%↑  &66s  &64.3      \\
50\% &  0.719 &  2.04\% ↓  &0.745 &1.19\%↓    &21m  &18.3    &18.63 &0.27\%↑  &20.23 &1.27\%↓  &69s  &53.2      \\  
\hline
\end{tabular}
\label{tab:SlimCodeResult}
\end{table}

\begin{table}[t]
\setlength{\tabcolsep}{3pt}
\centering
\small
\caption{Results of \underline{DietCode} for CodeBERT and CodeT5 on code search and summarization.(10\%-50\%:removing 10\%-50\% tokens for each snippet, R-M: Reduce MRR, BLEU:BLEU-4 values, R-B:Reduced BLEU-4 values, Time: the time for removing tokens in code snippet dose not include weight retrieval.)} 
%(DietCode provides only the result at a rate of 40\%.)}
\vspace{-4pt}
\begin{tabular}{l|cc|cc|c|cc|cc|c}
\hline
 \multirow{3}{*}{\textbf{Ratio}}& \multicolumn{5}{c|}{\textbf{Code Search}} & \multicolumn{5}{c}{\textbf{Code Summarization}} \\
\hline
 & \multicolumn{2}{c|}{\textbf{CodeBERT}} & \multicolumn{2}{c|}{\textbf{CodeT5}} & \multicolumn{1}{c|}{\textbf{Pruning}} & \multicolumn{2}{c|}{\textbf{CodeBERT}} & \multicolumn{2}{c|}{\textbf{CodeT5}} & \multicolumn{1}{c}{\textbf{Pruning}}\\
 & MRR & R-M &  MRR & R-M & Time & BLUE & R-B  & BLUE & R-B & Time \\
\hline
Base & 0.743 &0\%--     &0.754  &0.00\%--  &N/A     &18.58 &0\%--     &20.49  &0.00\%--  &N/A    \\
10\% & 0.702 &4.36\%↓  &0.730  &3.18\%↓   &9h24m   &17.68 &4.84\%↓  &19.68  &3.90\%↓   &1h40m  \\
20\% & 0.686 &6.54\%↓  &0.718  &4.77\%↓   &8h28m   &17.94 &3.44\%↓  &19.77  &3.51\%↓   &1h30m  \\
30\% & 0.693 &5.59\%↓  &0.714  &5.31\%↓   &7h37m   &17.73 &4.57\%↓  &19.68  &3.95\%↓   &1h19m  \\
40\% & 0.679 &7.49\%↓  &0.707  &6.21\%↓   &6h45m   &17.53 &5.65\%↓  &19.42  &5.22\%↓   &1h11m  \\
50\% & 0.651 &11.31\%↓ &0.676  &10.34\%↓  &5h59m   &17.67 &4.90\%↓  &19.22  &6.20\%↓   &1h02m  \\
\hline
\end{tabular}
\label{tab:DietCodeResult}
\end{table}

\subsubsection{Accuracy Comparison}

Table~\ref{tab:SlimCodeResult} and Table~\ref{tab:DietCodeResult} show that \tool outperforms the state-of-the-art \toolbase \cite{dietcode-fse22} in all experiments for code search and summarizaton tasks. 
%On average, the improvements on code search and summarization are 5.97\% and 4.23\%, respectively.} 
Specifically, \tool can improve \toolbase by 6.48\% and 4.27\% in terms of MRR (i.e., code search) and BLEU score (i.e., code summarization) for CodeBert, and by 5.47\% and 4.19\% in terms of MRR and BLEU score for CodeT5, respectively.
%When used with CodeBERT, {\tool} shows the highest and lowest improvements in code search and summarization, with values of \textcolor{blue}{9.46\%} and \textcolor{blue}{1.91\%}, respectively.
%\david{fill up xx\% in the above para}

Note that there are two interesting observations: 1) Intuitively, the performance of the removed code for downstream tasks might not be better than that for the original code, however, {\tool} can achieve better results. 2) Usually, the results of the code with a higher $SimplificationRatio$ should be worse. But, in some cases, the opposite is true for both methods. 
For instance, for CodeBERT in code search dataset, the performance of $SimplificationRatio = 30\%$ is better than that of $SimplificationRatio = 20\%$.

The reason is that
%the quality of code snippets determines the performance of downstream tasks, i.e., 
if the code contains more useful information, the downstream tasks will have better results.
In fact, not all code snippets can be fully inserted into models, as the maximum lengths for original codes are limited to 200 and 256 characters for code search and summarization. Therefore, even if some tokens are removed (e.g., 50\%) and the maximum lengths are reduced (e.g., $maxLength$=100), tokens towards the end of the code snippets may still be included in the input.
If the newly input tokens provide more information than the removed tokens, then the two interesting observations mentioned above may occur.

Actually, {\tool} often removes low-quality tokens (e.g., symbols) while allowing high-quality tokens to enter the model input. For instance, as reducing by 20\% tokens from a total of 200 tokens, the remaining top 160 tokens might still contain many symbol tokens. Consequently, more informative tokens sometimes do not make it into the top 160. However, when reducing by 30\% tokens, the additional 10\% for removal could contain more symbols. Thus, the remaining top 140 tokens could contain less symbol tokens and more important tokens could enter the top 140. Consequently, the result of $SimplificationRatio$=30\% could be better than that of $SimplificationRatio$=20\%.

\subsubsection{Percentage Distribution of Removed Tokens based on Token Categories}

\begin{figure}[t]
\centering
\stackengine{0pt}
{\kern7cm\includegraphics[height=4.1cm]{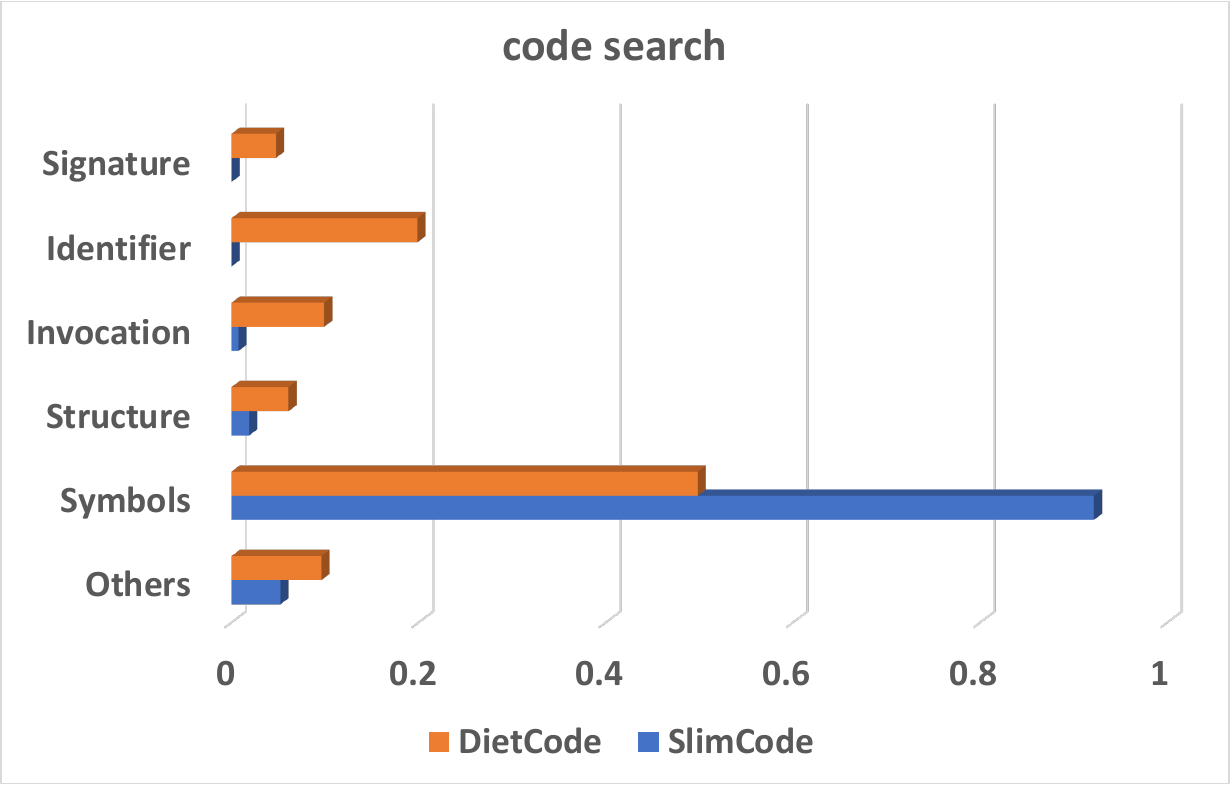}}
{\includegraphics[height=4.1cm]{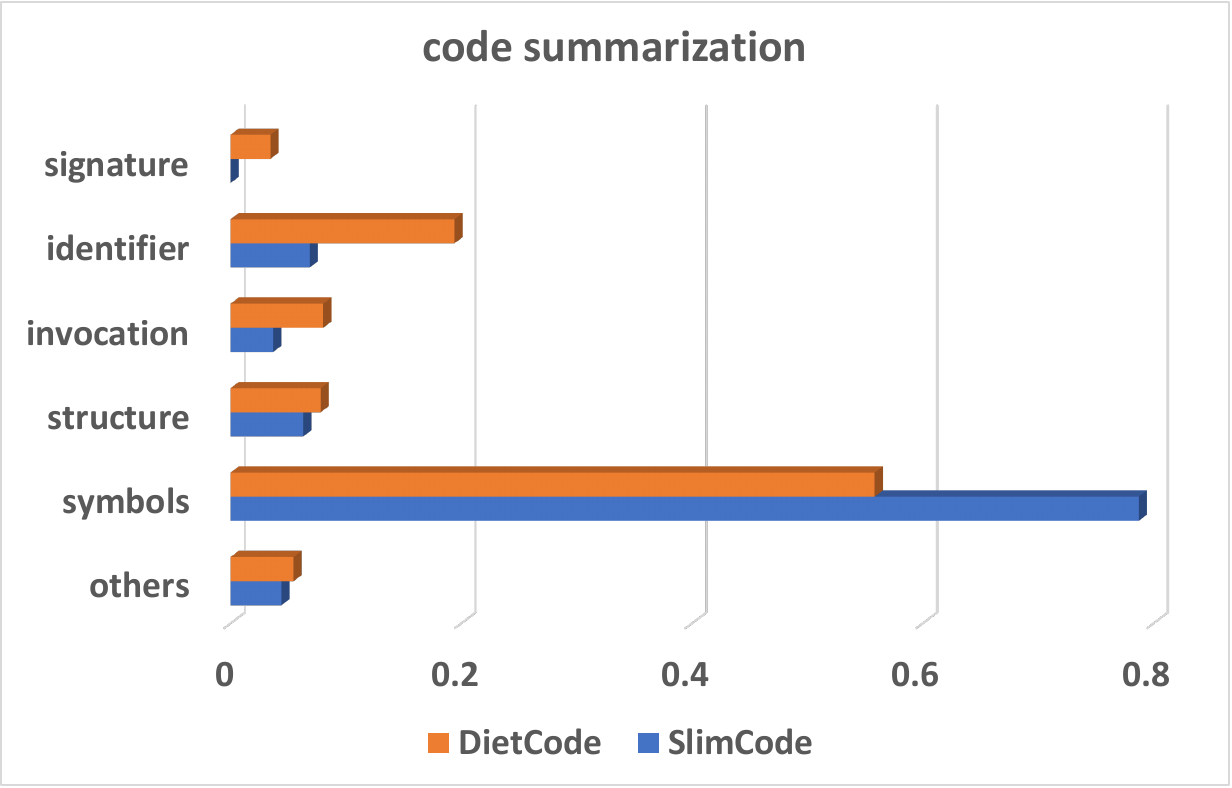}}
{O}{l}{F}{F}{L}
\vspace{-3pt}
\caption{The percentage distribution of removed tokens based on token categories.}
\label{fig:distribution}
\end{figure}

To further compare {\tool} and DietCode, we present the percentage distribution of removed tokens by token categories (Fig.~\ref{fig:distribution}). The figure illustrates that {\tool} selects tokens to be removed based on their ranking scores in both code search and code summarization (Algorithm~\ref{alg:slimcode}). 
Notably, removed tokens from the categories of symbol and other tokens can account for over 95\% of the total tokens removed. 
Meanwhile, {\tool} removes much fewer tokens from method signatures and identifiers compared to DietCode.
In contrast, while {\em DietCode also removes many symbol tokens that account for 50\% of the total tokens removed, it removes many high-ranking tokens (or partially removed after tokenization), especially identifiers}. 
As a result, much important semantic information is lost, leading to a decrease in downstream task performance.

\begin{comment}
\begin{figure*}[t]
  \centering
  \includegraphics[width=\textwidth,height=8cm]{figures/distribution_search.pdf}
  \caption{The percentage distribution of removed tokens based on token categories.}
  \label{fig:distribution}
  \vspace{-10pt}
\end{figure*}
\end{comment}

\begin{comment}
    \begin{figure}
    \begin{subfigure}
        \centering
        \includegraphics[width=\textwidth,height=8cm]{figures/distribution_search.pdf}
        \caption{The percentage distribution of removed tokens on code search.}
    \end{subfigure}
    \begin{subfigure}
        \centering
        \includegraphics[width=\textwidth,height=8cm]{figures/distribution_summarization.pdf}
        \caption{The percentage distribution of removed tokens on code summarization.}
    \end{subfigure}
\end{figure}
\end{comment}

%Analysis: 1) distrubution; 2) weights.
%+ demonstration example

%An interesting observation.

\begin{comment}
\begin{enumerate}
    \item The performance of our $SlimCode$ is comparable with the one of state-of-the-art DietCode and even slightly better at the simplification ratio of 40\%. 
    \item When the simplification ratio is less than 30\%, there is not much difference in performance. However, when it is greater than 30\%, our method's performance deteriorates quickly.    
\end{enumerate}
\end{comment}

%As the reduction ratio increases, our method's performance deteriorates. The main reason is that the performance remains relatively stable before the removal of symbol tokens, but deteriorates rapidly after the removal of other category tokens.

\subsubsection{Time Efficiency}

{\tool} can be up to 133 times faster than DietCode, which needs to run the model on the training dataset to obtain the weights of all tokens, and then remove tokens based on their weights.
Table~\ref{tab:SlimCodeResult} and Table~\ref{tab:DietCodeResult} show the running times of both token-removal algorithms on the training datasets, as well as the time efficiency improvement of {\tool} over Dietcode. 
From the tables, we can see that DietCode needs 1 hour to 10 hours to remove tokens with different $SimplificationRatio$s, and higher $SimplificationRatio$ means less pruning time. 
Notice that the pruning time of DietCode does not include the time for weight retrieval from a model. 

Compared to {\tool}, DietCode requires an additional step of selecting code statements, which is converted into a knapsack problem. This step is implemented by dynamic programming, and has a time complexity of $O(|D_t| \times n_s \times (1-SimplifiedRatio) \times (L+ml))$, where $|D_t|$ is the size of the training set, $n_s$ and $(1-SimplifiedRatio) \times (L+ml)$ are the average number of total statements and the number of candidate tokens to retain in code snippets, and $ml$ is the average length of the longest sentence in each code snippet.
The complexity of {\tool} is $O(|D_t| \times n_t)$ according to Algorithm~\ref{alg:slimcode}. Thus, even without considering other steps of DietCode, {\tool} is at least $n_s$ times faster than DietCode since usually $(1-SimplifiedRatio) \times (L+ml) > n_t$.
For example, in code search with $|D|=0.9M$, $n_s=17$, $SimplifiedRatio=10\%$, $L=200$, $n_t=112$, and $ml=20$, the ratio of $\frac{|D_t| \times n_s \times (1-SimplifiedRatio) \times (L+ml)}{|D_t|\times n_t}$ is 30. This is close to the experimental result of 32 times.

Moreover, the performance of {\tool} on code summarization can improve by up to 133 times, as the average length of code snippets in the summarization dataset has been reduced to 100. This means that more snippets do not need to be pruned with the given target length, and lines 6-10 of Algorithm~\ref{alg:slimcode} do not need to be executed. However, DietCode does not benefit from this change.

We also observe that DietCode using CodeBERT and CodeT5 generate two distinct lists of top 1,000 tokens using the learned weights from two models, with a low Spearman Correlation Coefficient value, i.e, 0.43. %This finding suggests that the token weights can vary between different models due to varying attention patterns and training datasets.
This finding suggests that the token weights from CodeBERT and CodeT5 are quite different. The reason could be that the pre-training loss functions, pre-training data, network structures, or tokenizers of CodeBERT and CodeT5 are not the same. Any of these differences can lead to changes in the weight of each token. Thus, {\em DietCode is model dependent, as its token weights learned from CodeBERT cannot be directly applied to CodeT5, and vise versa}.

%\textcolor{blue}{Zhang et al.~\cite{dietcode-fse22} 
%provide a list of low-rated tokens directly, which can be used to remove low-weight tokens for both models. 

We replicated the training processes to obtain the weights for all tokens in both models (they needs at least another 21 hours). We found that the Spearman Correlation Coefficient between the two lists of the top 1,000 tokens with the highest weights from the two models is low. This finding suggests that the token weights can vary between different models due to varying attention patterns and training datasets.

\intuition{
{\bf RQ5 Takeaway}:  
Overall, {\tool} has outperformed DietCode in code search and summarization by up to 9.46\% and 5.15\%, respectively. Importantly, {\tool} is much more efficient, being up to 133 times faster than DietCode. Importantly, {\tool} is model-agnostic, while DietCode is not as its attention scores for tokens learned from one model cannot be directly applied to another.
}

%\section{Do our findings hold for pre-train--prompt--prediction paradigm?}

\section{Usage of {\tool} for GPT-4 with zero-shot-learning}

\subsection{Analysis Procedure}

In this study, 
our goal is to evaluate whether the previous results and conclusions are applicable to the pre-train, prompt, and prediction paradigm. We replicate the analysis procedures in the above RQ1-4 with GPT-4. As GPT-4 only provides a Web API, we used the selenium package to simulate a browser and communicate with GPT-4. We conducted code search and code summarization tasks using two prompts: 1) for code search, {\em ``Please check whether the code snippet is semantically consistent with the given $text$. Please analyze step-by-step first, and then answer in the following format."}, and 2) for code summarization, {\em ``Write a short sentence to describe the function of the code snippet. Answer in the following format."} Finally, we analyzed the results of GPT-4 to calculate the number of total tokens, precision, and BLEU-4.

\subsection{Empirical Results of Replicating Previous Experiments on GPT-4}\label{chatgptresults}

\begin{table}[t]
\small
\centering
\caption{Results of removal methods on GPT-4 for \underline{code search}. (IT:Input Tokens, R-IT:Reduced Input Tokens in \%, OT: Output Token, R-OT:Reduced Output Tokens in \%; TT:Total Tokens, R-P:Reduced Precision in \%)}
\vspace{-6pt}
\resizebox{\linewidth}{!}{
\begin{tabular}{l|cc|cc|cc|cc}
\hline
\textbf{Removal method} & \textbf{IT} & \textbf{R-IT} & \textbf{OT} & \textbf{R-OT} & \textbf{TT} & \textbf{R-TT} & \textbf{Precision} & \textbf{R-P} \\
\hline
Base              & 69820 & 0.00\%     & 44584  & 0.00\%   & 114404  & 0.00\%--   & 0.85 & 0.00\% \\
\hline
Random(10\%)      & 65144 &  6.70\%↓   & 42352  &  5.01\%↓   & 109728  &  4.09\%↓     & 0.74 & 12.94\%↓ \\
Random(20\%)      & 60275 &  13.69\%↓  & 38844  &  12.87\%↓  & 99119   &  13.37\%↓    & 0.68 & 20.00\%↓ \\
Random(30\%)      & 55411 &  20.64\%↓  & 38204  &  14.31\%↓  & 93615   &  18.17\%↓    & 0.68 & 20.00\%↓ \\
Random(40\%)      & 50548 &  27.60\%↓  & 36852  &  17.34\%↓  & 87400   &  23.60\%↓    & 0.64 & 24.71\%↓ \\
Random(50\%)      & 45649 &  34.62\%↓  & 33472  &  24.92\%↓  & 79121   &  30.84\%↓    & 0.63 & 25.88\%↓ \\
\hline
Identifier        & 60976 &  12.67\%↓  & 39637  &  11.10\%↓  & 10063   &  12.10\%↓    & 0.66 & 22.35\%↓ \\
Symbol tokens     & 43276 &  38.02\%↓  & 39049  &  12.42\%↓  & 82325   &  28.04\%↓    & 0.78 & 8.24\%↓ \\
Control structures & 60032 &  14.02\%↓  & 41257  &  7.46\%↓   & 101289  &  11.46\%↓    & 0.65 & 23.53\%↓ \\
Method invocations & 49824 &  28.63\%↓  & 41348  &  7.26\%↓   & 132520  &  15.83\%↓    & 0.70 & 17.65\%↓ \\
Method signature  & 64925 &  7.01\%↓   & 39324  &  11.80\%↓  & 104249  &  8.88\%↓     & 0.70 & 17.65\%↓ \\
PDG               & 52360 &  25.01\%↓  & 40459  &  9.25\%↑   & 92819   &  18.87\%↑    & 0.71 & 16.47\%↓ \\
\hline
DietCode(10\%)    & 64776 & 7.22\%↓    & 40328  & 9.55\%↓    & 105104  & 13.94\%↓     & 0.65 & 23.53\%↓ \\
DietCode(20\%)    & 59941 & 14.15\%↓   & 36753  & 17.56\%↓   & 96694   & 19.93\%↓     & 0.64 & 24.71\%↓ \\
DietCode(30\%)    & 55036 & 21.17\%↓   & 39300  & 11.85\%↓   & 94336   & 21.26\%↓     & 0.62 & 27.06\%↓ \\
DietCode(40\%)    & 50224 & 28.06\%↓   & 37061  & 16.87\%↓   & 87285   & 25.62\%↓     & 0.57 & 32.94\%↓ \\
DietCode(50\%)    & 45388 & 34.99\%↓   & 35364  & 20.68\%↓   & 80752   & 30.23\%↓     & 0.56 & 34.11\%↓ \\
\hline
{\tool}(10\%)     & 64776 & 7.22\%↓    & 40988  & 8.06\%↓    & 105769  & 7.55\%↓      & 0.78 & 8.24\%↓ \\
{\tool}(20\%)     & 59941 & 14.15\%↓   & 39500  & 11.40\%↓   & 99441   & 13.08\%↓     & 0.71 & 16.47\%↓ \\
{\tool}(30\%)     & 55036 & 21.17\%↓   & 38224  & 14.27\%↓   & 93260   & 18.48\%↓     & 0.74 & 12.94\%↓ \\
{\tool}(40\%)     & 50224 & 28.06\%↓   & 38616  & 13.39\%↓   & 88840   & 22.34\%↓     & 0.74 & 12.94\%↓ \\
{\tool}(50\%)     & 45388 & 34.99\%↓   & 37879  & 15.05\%↓   & 83267   & 27.22\%↓     & 0.87 & 2.35\%↑ \\
\hline
\end{tabular}
}
\label{tab:gpt_code_search}
%\vspace{-5pt}
\end{table}
%\vspace{-5pt}

Tables~\ref{tab:gpt_code_search} and~\ref{tab:gpt_code_summarization} show the results of various code simplification methods on code search and code summarization respectively. 
Note that, here we use the total number of tokens (the input tokens and output tokens) to replace the prediction time, mainly because 1) the fee charged for the usage of computing resource of GPT-4 is based on the total number of tokens~\cite{openai-pricing}; 2) GPT-4 only provides a web interface, and the actual prediction time is affected by multiple factors, which is usually difficult to measure accurately. 
Moreover, the input tokens not only include the code snippet, but also the prompts.
%and the system prompt automatically generated by GPT-4~\cite{DBLP:journals/corr/abs-2303-08774}. 
From the Tables, we have the following empirical findings:

\subsubsection{Finding 1}

As the simplified ratio increases, the total number of tokens and metrics decrease for both tasks in GPT-4.
This empirical finding substantiates that code simplification leads to a reduction in computational resources required for Prompt-based Pattern Learning Model (PPLM). This can be attributed to the fact that GPT-4 requires the generation of a significant amount of analytical content by following a sequence of statements through reasoning, as opposed to solely producing binary responses. It demonstrates that in the absence of generating analytical text, the classification performance of GPT-4 is notably deficient, a characteristic linked to its pre-training methodology~\cite{DBLP:journals/corr/abs-2303-08774}. 
As the code becomes more incomplete, the analytical content of GPT-4 on the remaining code also decreases. Consequently, the total number of generated tokens and metrics decreases as well.
Furthermore, GPT-4 lacks the capability to produce code descriptions closely resembling the ground truth, as it cannot undergo fine-tuning. In other words, it lacks exposure to the specific code description corresponding to each code snippet. The overlap of words between the generated description and the ground truth is considerably lower when compared to CodeBERT and CodeT5. Thus, the BLEU-4 scores exhibit a marked reduction relative to their preceding values and demonstrate a proximity to one another.

\subsubsection{Finding 2}

For code search, simplified code based on different token categories still shows different performance based on the GPT-4 model. Compared to CodeBERT and CodeT5, their impacts remain almost unchanged except that tokens in control structures and invocations have greater impacts. In the case of code summarization, the results across different categories tend to have a similar appearance, which differs from previous cases. 

We find that in code search, tokens within the method structures and invocations exhibit greater significance compared to tokens within the method signatures. This disparity may be attributed to their heightened influence on GPT-4's sequential reasoning process. 
Tokens within the method structures and invocations directly encapsulate the code's execution process, providing richer details about the code.
Without the details, there cannot be a proper sequential reasoning process, and the quality of classification and text generation of GPT-4 will be severely compromised.
Conversely, tokens within the method signature offer a broader overview of the code's functionality. 
%\textcolor{blue}{In code summarization, the BLEU-4 scores are directly proportional to the degree of word overlap between the generated description and the ground truth. Conversely, there is an inverse relationship with the number of outputs. Since both the overlap and the number of outputs do not change significantly, the difference in BLEU-4 values for category tokens is not substantial.}
%For code summarization, there is a direct proportionality between BLEU-4 scores and the degree of the overlap of words between the generated description and the ground truth, while an inverse relationship exists with the number of outputs. 
%Because both the overlap and the number of outputs do not change significantly, the difference in BLEU-4 values of category tokens is not substantial.
%This relationship is further influenced by token categories and the applied simplified ratio. 
It is evident that tokens involved in method invocations exert the most pronounced impact, while those associated with PDG have the smallest effect. This is likely because disrupting the sequence of method invocations significantly influences the generation of code descriptions in accordance with the code's order. Additionally, tokens in method signatures, identifiers, and control structures exhibit relatively substantial effects on BLEU-4 scores. Despite symbol tokens having a BLEU-4 score of only 5.22, their simplified ratio stands at 42\%. Consequently, eliminating symbol tokens remains the most effective strategy to enhance code summarization.

\begin{table}[t]
\small
\centering
\caption{Results of various removal methods on GPT-4 for \underline{code summarization}.(IT:Input Tokens, R-IT:Reduced Input Tokens in \%, OT: Output Token, R-OT:Reduced Input Tokens in \%; TT:Total Tokens, R-B:Reduce BLEU-4 in \%)}
%\vspace{-6pt}
\resizebox{\linewidth}{!}{
\begin{tabular}{l|cc|cc|cc|cc}
\hline
\textbf{Removal method} & \textbf{IT} & \textbf{R-IT} & \textbf{OT} & \textbf{R-OT} & \textbf{TT} & \textbf{R-TT} & \textbf{BLEU-4} & \textbf{R-B} \\
    \hline
    
Base & 48037 & 0.00\%- & 15348 & 0.00\%- & 63385 & 0.00\%- & 5.50 & 0.00\%- \\
\hline
Random(10\%) & 44361  & 7.65\%↓   & 16249 & 5.86\%↑ & 60610 & 4.38\%↓  & 5.36 & 2.55\%↓ \\
Random(20\%) & 40509  & 15.67\%↓  & 15532 & 1.20\%↑ & 56041 & 11.59\%↓ & 5.49 & 5.37\%↓ \\
Random(30\%) & 36680  & 23.64\%↓  & 15401 & 0.34\%↑ & 52081 & 17.83\%↓ & 5.43 & 1.27\%↓ \\
Random(40\%) & 32813  & 31.69\%↓  & 15424 & 0.50\%↑ & 48237 & 23.90\%↓ & 5.25 & 4.55\%↓ \\
Random(50\%) & 28912  & 39.81\%↓  & 14961 & 2.53\%↓ & 43873 & 30.78\%↓ & 4.96 & 8.15\%↓ \\
\hline
Identifier        & 41665   & 13.27\%↓ & 15392 & 0.29\%↑ & 57057 & 9.98\%↓  & 5.48  & 0.36\%↓ \\
Symbol tokens     & 27680   & 42.38\%↓ & 16688 & 8.73\%↑ & 44368 & 30.00\%↓ & 5.22  & 5.09\%↓ \\
Control structures & 42301   & 11.94\%↓ & 16024 & 4.40\%↑ & 58325 & 7.98\%↓  & 5.44  & 1.09\%↓ \\
Method invocations & 33168   & 30.95\%↓ & 15501 & 0.99\%↑ & 48669 & 23.22\%↓ & 4.80  & 12.72\%↓ \\
Method signature  & 43664   & 9.09\%↓  & 16313 & 6.28\%↑ & 59981 & 5.37\%↓  & 5.21  & 5.27\%↓ \\
PDG               & 42232   &12.08\%↓  & 15104 & 1.59\%↓ & 57336 & 9.54\%↓  & 5.78  & -5.09\%↑ \\
\hline
DietCode(10\%) & 43876 & 8.66\%↓  & 16080 & 4.77\%↑ & 59956 & 5.41\%↓  & 4.96 & 9.82\%↓ \\
DietCode(20\%) & 40105 & 16.51\%↓ & 14845 & 3.28\%↓ & 54950 & 13.31\%↓ & 5.27 & 4.18\%↓ \\
DietCode(30\%) & 36277  & 24.48\%↓ & 15076 & 1.77\%↓ & 51353 & 18.98\%↓ & 5.00 & 9.09\%↓ \\
DietCode(40\%) & 32484  & 32.38\%↓ & 15100 & 1.62\%↓ & 47584 & 24.93\%↓ & 4.66 & 15.27\%↓ \\
DietCode(50\%) & 28724  & 40.20\%↓ & 14521 & 5.39\%↓ & 43245 & 31.77\%↓ & 4.64 & 15.63\%↓ \\
\hline
{\tool}(10\%) & 44005  & 8.39\%↓  & 15868 & 3.62\%↑ & 59873 & 5.48\%↓  & 5.42  & 1.45\%↓ \\
{\tool}(20\%) & 40188  & 16.34\%↓ & 16544 & 7.79\%↑ & 56732 & 10.49\%↓ & 5.40  & 1.82\%↓ \\
{\tool}(30\%) & 36321  & 24.39\%↓ & 16092 & 4.85\%↑ & 52413 & 17.31\%↓ & 5.50  & 0.00\%- \\
{\tool}(40\%) & 32492  & 32.36\%↓ & 16593 & 8.11\%↑ & 49085 & 22.56\%↓ & 5.20  & 5.45\%↓ \\
{\tool}(50\%) & 28724  & 40.20\%↓ & 15848 & 3.26\%↑ & 44572 & 29.68\%↓ & 5.51  & 2.04\%↑ \\
\hline
%\vspace{-3pt}
\end{tabular}
}
\label{tab:gpt_code_summarization}
\end{table}

\subsubsection{{\tool} and DietCode applied on GPT-4}

On average, {\em {\tool} outperforms \toolbase with a 26\% improvement in precision for code search and a 10\% increase in BLEU-4 for code summarization}. In fact, the simplified code produced by {\tool} can even surpass the original code in performance.
It is evident that the performance enhancement achieved by {\tool} surpasses that of DietCode in comparison to the improvements by CodeBERT and CodeT5. This discrepancy arises primarily from DietCode's approach of simplifying code through statement deletion, which poses a substantial challenge for GPT-4. Given GPT-4's reliance on sequential code analysis followed by downstream tasks, this method may lead to GPT-4 failing to accurately infer the content of the removed statements. Conversely, {\em {\tool}'s strategy of removing symbol tokens proves more advantageous for GPT-4 in "recovering" the deleted tokens based on its pre-trained knowledge}. This assertion is supported by results obtained through a random removal method. 

Notably, an intriguing phenomenon emerges wherein {\em {\tool} with a $Simplified Ratio$ of 50\% outperforms the original code on both tasks}. With a reduction ratio of 50\%, all symbol tokens, along with certain tokens within control structures and method invocations, are removed. At this, GPT-4 encounters difficulty in reconstructing specific code details. Consequently, it leans more heavily on high-level semantic information gleaned from method signatures and identifiers, in conjunction with its pre-training knowledge, to execute downstream tasks. This adjustment may in fact enhance the efficacy of these downstream tasks. One conceivable explanation lies in the fact that when the original code contains an abundance of specific code details, GPT-4 tends to rely heavily on these details to perform downstream tasks. However, if these details are "inaccurate" (e.g., due to illogical variable or method names), GPT-4 is more prone to identify them as "matching contradictions" or "generating partially unreasonable expressions," resulting in a decrease in effectiveness.

If we remove 50\% tokens by using {\tool} (the last row of Table~\ref{tab:gpt_code_search}), the precision of simplified code can be even better than that of the original code, with the improvement of precision about 2.35\%. Additionally, the total number of tokens is reduced by 27.22\%, which leads to cost saving because 
according to OpenAI's pricing policy~\cite{openai-pricing}, the cost is proportional to the number of tokens. 
Removing symbol tokens saves the cost of invoking GPT-4 by 24\% here since the total cost of 400 samples just is 9.54 dollars.
As the use of GPT-4 and other PLLMs becomes more prevalent, companies are likely to use them to solve thousands of other tasks. Our study can yield better results, save much time, and reduce more expenses. The same trend is observed for code summarization.

\intuition{
{\bf RQ6 Takeaway}: 
Experimental validation of the GPT-4 model confirmed that all findings from the previous research questions hold for the pre-training, prompt, and prediction paradigm. Importantly, {\tool} can yield better results, save inference time, and reduce expenses for smaller input sizes.
}

\section{Discussions and Potential uses of {\tool}}

%{\color{red}{

%we can talk the following points:

%1. possible usage scenarios of using our findings in pre-train small models, like codeT5 and codebert: including who are the practitioners to use this?, how should they use our results and tools

%2. possible usage scenarios of using our findings in chatgpt-like models (RQ6)

%3. lessons learned during the process of our empirical study: lesson 1, xxxx, lesson 2 xxxx

%4. we can also talk about a little bit the future work what we want to do more.

%}}

%Here are some of the key implications and potential uses of {\tool}:

\subsection{Computational Efficiency Improvement} 

The computational complexity of code-oriented LLMs has been a concern, particularly as the length of input code sequences increases. Our work addresses this challenge by simplifying input code before feeding it to the LLMs. The linear-like relationship between code reduction and training time savings is a significant finding. This implies that by applying {\tool}, developers can substantially reduce the time required for model training and inference, making LLMs more practical and efficient.

\subsection{Generalizability Across Models and Tasks}

We showed that the impact of categorized tokens on code simplification is both model-agnostic and task-specific. This finding is crucial since it highlights that {\tool} can be applied across different LLM architectures such as CodeBERT, CodeT5, and GPT-4. This generalizability needs more experiments to ensures that the benefits of {\tool} can be realized across LLMs, making it a versatile solution for code simplification. More experiments are also needed with other tasks.

\subsection{Reduced Dependency on Attention Patterns}
 
{\tool}, which is not dependent on attention patterns, is a more robust and stable method for code simplification. This is important when the LLM is trained on different datasets, as the outcomes might vary due to varying attention patterns.

\subsection{Cost Savings in API Usages to LLMs}

Our study shows that implementing {\tool} can result in a significant reduction in the cost of invoking LLMs through API queries, specifically GPT-4. This cost-saving potential has practical implications for organizations and developers that heavily rely on LLMs for code-related tasks. By reducing the number of API queries needed, organizations can effectively manage their resources while still benefiting from the power of LLMs.

\section{Related work}

\subsubsection*{Pre-trained Models in Software Engineering and its Understanding} Pre-trained technologies have achieved a remarkable success in Natural Language Processing (NLP)~\cite{devlin2018bert,yang2019xlnet,liu2019roberta} and pre-trained models (PTM) associated with programming languages~\cite{codebert-emnlp20,kanade2019pre,guo2020graphcodebert,karampatsis2020scelmo,lewis2019bart,guo2022unixcoder,li2022codereviewer,wang2019extracting, li2021context, li2021fault} also make a great process of code intelligence, including code repair~\cite{li2020dlfix,li2022dear}, code generation~\cite{clement2020pymt5,wang2023codet5+}, defect detection~\cite{ni2022best,wang2023deepvd,ni2022defect,li2021vulnerability}, code summarization~\cite{liu2019roberta,ahmed2022multilingual,codebert-emnlp20,jiang2021treebert}, code search~\cite{wang2021syncobert,liu2022deeplearning}, etc.
%s
Due to pre-trained models' powerful ability, many works~\cite{ahmad2021unified,mastropaolo2021studying,paltenghi2021thinking,rogers2021primer,karmakar2021pre,DBLP:conf/iclr/AllamanisBK18} investigated the mechanisms of PLM for code.
%For example, Karmakar and Robbes~\cite{karmakar2021pre} empirically adopted four probing tasks on PTMs to investigate whether PTMs can learn different aspects of source code (e.g., syntactic, structural, surface-level, and semantic information).
Zhang et al.~\cite{dietcode-fse22} focused on the specific knowledge (e.g., critical tokens/statements) learned by PLMs. Wan et al.~\cite{wan2022they} performed a structural analysis of PLMs for source code and found that the Transformer attention can capture high-level structural information. Ahmed and Devanbu~\cite{ahmed2022multilingual} studied the impact of identifiers and that of non-identifiers. 

%Bui et al.~\cite{bui2019autofocus} proposed Autofocus to reveal the most relevant part of the code to programmers by measuring 
%the relevance of statements using attention weights from a GGNN~\cite{DBLP:conf/iclr/AllamanisBK18}.

\subsubsection*{Program Simplification} Several state-of-the-art approaches are proposed, including 
SIVAND~\cite{rabin2021understanding} and P2IM~\cite{suneja2021probing}.
These approaches are usually based on the delta debugging prototype~\cite{zeller2002simplifying}, which treats a code snippet and an auxiliary deep learning model (e.g., code2vec) as the input.
The model splits the code snippet into fragments and each fragment is subsequently treated as the input of the neural network model for performing testing tasks (e.g., method name prediction or misused
variables detection).
%Meanwhile, a fragment will be further split if it achieves a satisfying score. 
%Such a process continues until the performance of the subset fails to obtain a satisfying score.
%Eventually, the smallest code snippet that satisfies the objective of the deep learning model will be obtained.
These methods are computationally inefficient (e.g., hundreds of hours for SIVAND to process 10K functions) since they need to run a deep learning model and evaluate the performance at each iteration.
To solve that, Zhang et al.~\cite{dietcode-fse22} proposed DietCode to improve further running efficiency (e.g., two minutes for DietCode to process 10K functions).

\section{Conclusion}

In summary, we propose {\tool}, a model-agnostic solution to code simplification that has far-reaching implications in terms of computational efficiency, generalizability, cost savings, and broader applicability within the field of software engineering (SE). The departure point of {\tool} is the exploration of the nature of code tokens and their impacts on the performance of LLMs. It not only addresses current challenges but also opens up new avenues for improving various SE tasks that rely on LLMs. The empirical results show that {\tool} can improve over DietCode by  9.46\% and 5.15\% in terms of MRR and BLEU score on code search and summarization, respectively. More importantly, \tool is 133 times faster than \toolbase.  Additionally, {\tool} can reduce the cost of invoking GPT-4 by up to 24\% per API query, while still producing comparable results to those with the original code. 
We also call for a new direction on code-based, model-agnostic code simplification solutions to further empower LLMs in many other software engineering tasks. 

%This paper calls for a new direction on code-based, model-agnostic code simplification solutions to further empower LLMs.

%In summary, our research offers a model-agnostic solution to code simplification that has far-reaching implications in terms of computational efficiency, generalizability, cost savings, and broader applicability within the field of software engineering. {\tool} not only addresses current challenges but also opens up new avenues for improving various SE tasks that rely on LLMs. The empirical results show that \tool can improve the state-of-the-art technique \toolbase by {\bf 9.46\%} and {\bf 5.15\%} in terms of MRR and BLEU score on code search and summarization, respectively. More importantly, \tool is {\bf \underline{133 times}} faster than \toolbase.  Additionally, {\tool} can reduce the cost of invoking GPT-4 by up to {\bf 24\%} per API query, while still producing comparable results to the original codes.This paper calls for a new direction on code-based, model-agnostic code simplification solutions to further empower LLMs.

\section{Data Availability}
Our code and experimental code are publicly available at https://github.com/gksajy/slimcode.

%\begin{enumerate}
%    \item There are two types of attention used in text classification and generation tasks. In classification (code search), one is the encoder attention and the other is the classification attention. In text generation, one is the encoder attention and the other is the decoder attention (encoder-decoder attention).
%    \item The encoder attention in pre-train include bimodal/unimodal cross and sequence attentions that enable each token to learn a corresponding semantic representation. Tokens with strong correlations share similar embeddings. The classification attention is us. 
%    \item To simplify code snippets for text classification and generation tasks, we should follow the classification and decoder attentions, not the encoder attention, as in DietCode.
%\end{enumerate}

% or


%\appendices

%\section{}

%\section*{Acknowledgment}

\section*{Acknowledgments}
The author Tien N. Nguyen is supported in part by the US National Science Foundation (NSF) grant CNS-2120386 and
the National Security Agency (NSA) grant NCAE-C-002-2021.


%\ifCLASSOPTIONcaptionsoff
% \newpage
%\fi

\balance

\bibliographystyle{ACM-Reference-Format}

\bibliography{references}

\end{document}